\pdfoutput=1

\documentclass[authoryear,preprint,12pt]{elsarticle}

\usepackage[utf8]{inputenc}
\usepackage{graphicx}
\usepackage{geometry}
\usepackage{pdflscape}   
\usepackage{longtable}
\usepackage{booktabs}
\usepackage{array}       
\usepackage{ragged2e}    
\usepackage{changepage}
\usepackage{makecell}

\usepackage{amsmath}
\usepackage{amssymb}
\usepackage{appendix}
\usepackage{tabularx}

\usepackage{lineno}
\usepackage{url}
\usepackage[hidelinks]{hyperref}

\usepackage{siunitx}
\journal{Icarus}
\usepackage{textcomp}
\DeclareUnicodeCharacter{2032}{\textprime}

\begin{document}
\begin{frontmatter}

\title{A Search for Hydroacoustic Signals from Bolides\tnoteref{t1}}
\tnotetext[t1]{Accepted for publication in \emph{Icarus} on April 14, 2026. This is an arXiv preprint (author accepted manuscript).}

\author[UWO,WS]{Peter Brown\corref{cor1}}\ead{pbrown@uwo.ca}

\author[UWO,WS]{Luke McFadden}\ead{lmcfadd6@uwo.ca}

\author[NRCAN]{David McCormack}\ead{david.mccormack@nrcan-rncan.gc.ca}

\author[NRCAN]{Mareike Adams}\ead{mareike.adams@nrcan-rncan.gc.ca}

\author[UWO,WS]{Denis Vida}\ead{dvida@uwo.ca}

\cortext[cor1]{Corresponding author}

\affiliation[UWO]{organization={Department of Physics and Astronomy, University of Western Ontario},
            addressline={1151 Richmond Street}, 
            city={London},
            postcode={N6A 3K7}, 
            state={Ontario},
            country={Canada}}

\affiliation[WS]{organization={Institute for Earth and Space Exploration, University of Western Ontario},
            addressline={Perth Drive}, 
            city={London},
            postcode={N6A 5B8}, 
            state={Ontario},
            country={Canada}}
            
\affiliation[NRCAN]{organization={Natural Resources Canada},
            addressline={2617 Anderson Road}, 
            city={Ottawa},
            postcode={K1A 0Y3}, 
            state={Ontario},
            country={Canada}}
            
\begin{abstract}
Airwaves from fireballs have been detected infrasonically and via seismo-acoustic coupling, but to date, there has not been a confirmed hydroacoustic detection of a fireball. Here we present a survey aimed at detecting hydroacoustic signals from fireballs  using the six hydrophone stations operated as part of the Comprehensive Test Ban Treaty Organisation (CTBTO) International Monitoring System. We identified 30 fireballs where propagation paths to stations exist. These included high energy fireballs (E $\geq$ 5 kT), those which occurred over favorable locations for coupling into the deep ocean as well as a selection of bolides close to CTBTO hydrophone stations. The largest of these impactors were $>$ 5 meters in diameter.  From theoretical and empirical considerations we show that direct hydroacoustic shock transmission is the most likely source mechanism, though large meteorites impacting the ocean surface from a fireball might be detectable in extreme cases.  
    We find one possible instance of a fireball occurring on Sep 2, 2003 off the coast of Alaska, where a linked hydroacoustic signal with the expected timing and backazimuth is detected. However, given the size of our survey and the random background rate of signals, this detection is statistically weak. 
    We conclude that hydroacoustic detection in the SOFAR channel of fireballs is very rare. Using our chosen set of signal processing parameters, assuming direct path H-phase signals, adopting a signal celerity range of 1.42-1.55 km/s we find no unambigous detections in 53 station-fireball pairs. Based on SOFAR-equivalent yields derived assuming the minimum detectable amplitude signal family association is representative of the noise background in our survey we estimate a conditional upper limit for fireball coupling efficiency of order 10$^{-10}$. A single well recorded airplane impact provides an empirical estimate for the energy coupling of surface ocean impacts to the SOFAR channel of 10$^{-4}$ for high velocity surface impacts.
\end{abstract}

\begin{highlights}
\item Survey of 30 bright fireballs for associated hydroacoustic signals
\item A 1 kT bolide on Sep 2, 2003 possibly detected
\item We set limits on fireball hydroacoustic coupling efficiency ($\eta$) to be $\leq$10$^{-10}$
\item Based on a well characterized aircraft oceanic impact, we find an energy transfer efficiency for surface ocean impacts into the SOFAR channel of 10$^{-4}$.

\end{highlights}


\end{frontmatter}

\section{Introduction}

Fireballs (or equivalently bolides) are the light, heat, and shock waves produced when a meteoroid of order decimeters or larger impact a planetary atmosphere. The fireball represents the final stage in a meteoroid's evolution in the inner solar system and may, in some cases, drop meteorites. Recovery of meteorites and knowledge of a fireball's pre-atmospheric orbit can be used to infer the original source region of the meteorite in the main asteroid belt \citep{Borovicka_2015}. 

The detection and characterization of fireballs is important for validation and refinement of planetary defence models \citep{Mathias_2017, Wheeler2018}. The most commonly used technology for fireball characterization are optical cameras, recorded from either ground-based \citep{Koten2019} or space-based systems \citep{Tagliaferri1994}. Rarely, radar recordings may also detect the ionization left behind from a fireball \citep{Brown2011}. Ground-based optical and radar measurements suffer from comparatively small atmospheric collection areas a limitation which can be overcome with space-based detections, but usually at the expense of lower measurement precision.

Several fireball detection techniques exploit the shock wave produced by the hypersonic passage of a large meteoroid. For more energetic bolides (associated with meter-class and larger objects) the shock effects may include gravity-wave perturbations in the ionosphere \citep{Yang_2014} and low frequency sound detectable over near global distances by infrasound arrays \citep{Silber2019}. 

Fireball shock waves decay to infrasound a short distance (of order tens of kilometers for large fireballs) from the meteor trajectory \citep{Revelle1976}. Infrasonic waves can be transmitted thousands of kilometers due in part to the low attenuation of atmospheric propagation of infrasound ($<$20 Hz)\citep{Sutherland2004}, making them ideal for global fireball detection \citep{Ens2012, Silber2025}. Long-range propagation is also aided by the presence of atmospheric ducts, wherein acoustic waves move horizontally through waveguides created by changes in sound speed with height. The acoustic wave energy becomes trapped between heights when the local speed of sound is lower than above and below the guide height \citep{Lepichon2019}.

\begin{figure}
  \includegraphics[width=\linewidth]{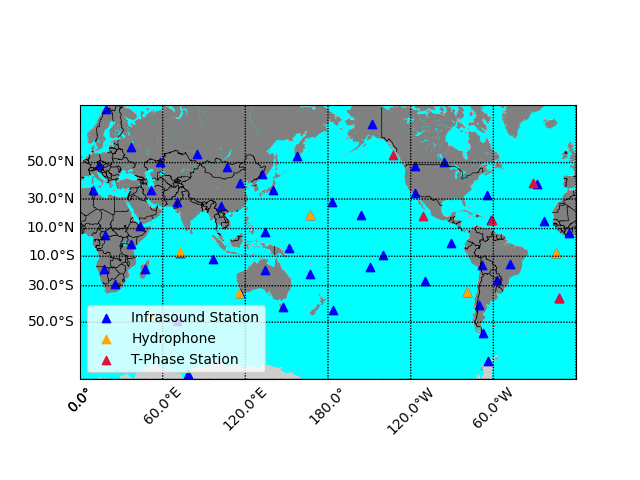}
  \caption{The location of infrasound, hydrophone, and T-phase stations (seismometers on islands where hydroacoustic signals are transmitted through the solid Earth to land) which form part of the International Monitoring System (IMS) operated by the Comprehensive Test Ban Treaty Organisation (CTBTO).}
  \label{ctbto_ims}
\end{figure}

The number of fireballs detected acoustically has increased substantially since the formation of the Comprehensive Nuclear-Test-Ban Treaty Organisation (CTBTO) in 1996. The CTBTO established seismic, infrasound, hydroacoustic, and radionuclide stations with the goal of global detection of any nuclear detonations (see Figure \ref{ctbto_ims})\citep{Lepichon2019} in the form of the International Monitoring System (IMS). As fireballs may produce aerial explosions of a similar energy to nuclear devices they are sometimes also referred to as airbursts in analogy with nuclear detonations \citep{Glasstone1977} and are a source of interest to CTBTO \citep{Silber2019}. Bolide airwave detections using seismic and infrasonic methods have been extensively documented \citep{Anglin1988, Cumming1989, Qamar1995, Tatum2000, pichon2002, Edwards2003, Langston2004, Ishihara2004, Pujol2005, Mcfadden2021, Silber2025}. Such prior works have demonstrated that airwaves can be used to estimate fireball source energies as well as for geolocation purposes and in special cases trajectory reconstruction.

When the infrasonic wave resulting from a meteor shock reaches the Earth's surface, seismo-acoustically coupled waves can be detected by seismometers \citep{Edwards2008}. The typical seismic coupling efficiency for atmospheric shocks is of order $10^{-2} - 10^{-7}$, depending on bolide energy  \citep{Edwards2008, Svetsov2019}. Nevertheless, airwaves from fireballs are commonly detected by seismometers on Earth \citep{Stich2022} and more recently on Mars \citep{Garcia2022}. 

Fireball shock waves are also expected to couple to the ocean. This should be more efficient compared to coupling with rock due to waters lower acoustic impedance \citep{Benmenahem1981}. An energy transmission of about 0.1\% (-29.5 dB) is found for a vertically incident plane wave \citep{Chapman1990} at the air-ocean boundary. However, the energy efficiency drops exponentially as the signal propagates as an evanescent wave at large depths \citep{Sawyers1968}. The transmission loss is also much higher when the receiver is located at large horizontal distances from the airborne source (i.e. the wavefront incidence angle is low). As large fireball airbursts tend to occur at great distances from sparsely installed hydrophones this suggests hydroacoustic detection of coupled fireball shocks will be challenging.

In contrast, underwater impulsive sources are much easier to detect.  Examples of studies which used hydroacoustic arrays from the CTBTO for detecting natural sources include those focused on volcanoes \citep{LeBras2023} and earthquakes \citep{Tolstoy2006}. Anthropogenic sources have also been well studied, including an underwater battery explosion \citep{Prior2010}, hydroacoustic detections of plane crashes into the water \citep{Metz2022, Kadri_2024}, and the implosion of the San Juan submarine which sank in 2017 \citep{Vergoz2021, Nielsen2021, Prario2023}.

Hydroacoustic detection of fireballs are non-existent to our knowledge in the literature. The challenge of hydroacoustic detection of atmospheric explosions is the relatively high background noise in the ocean compared to atmospheric noise \citep{Godin2008} coupled with the low transmission efficiency across the air-water interface to the depth of the Sound Fixing and Ranging (SOFAR) channel, (located at roughly 1 km in depth), an underwater analog to acoustic ducts in the atmosphere. As a result, there is limited open literature on the hydroacoustic signatures of airborne explosive/impulsive sources. We were only able to identity \cite{Kadri2017} as mentioning hydroacoustic signals from fireballs in a work focused on examination of signals potentially associated with the crash of the MH370 airliner. They localized a hydroacoustic signal which they suggested might be consistent with a meteorite impact in the Pacific. However, no corresponding bolide data was presented and this remains an unproven association.

Several theoretical works have examined the detection of airborne shock waves underwater \citep{Sparrow1997, Cheng1998}. In addition, detections of sonic booms \citep{Sohn2000} at shallow ($\leq$100 m) depths and above-water nuclear explosions \citep{Adushkin2004a} have previously been documented. 

\cite{Sawyers1968} developed a flat-interface model which was among the first to accurately predict that sonic booms penetrate the ocean as evanescent waves, with pressure decaying exponentially with depth. While his model was validated in several empirical studies \citep{Cook1967, Waters1971, Young1973} the focus in these works were on aircraft generated sonic booms and therefore were appropriate to small energy (and high frequency) airwaves. This is in contrast to the lower frequencies expected for energetic fireballs. Subsequent works have consistently shown that low-frequency components of the sonic boom (most germane to airbursts) propagate farther underwater, enabling their detection at shallow depths using hydrophones \citep{Sohn2000, DesharnaisChapman2002}. Coupling from water to air for signal wavelengths much larger than the source depth can also produce anomalously high transmission across the water interface as shown by \citet{Evers2014}, emphasizing the complexity of the energy transfer process.

More recently, studies incorporating real-world complexities, such as wavy ocean surfaces, bubble layers, and coastal sediments have provided insight into the changes in attenuation/coupling as a function of local ocean conditions. For example, both theoretical and empirical studies incorporating wavy surface effects have revealed that sonic booms interacting with surface wave trains produce non-evanescent wave components, reducing attenuation at depth and generally providing better air-water coupling \citep{Cheng1998}.  Similarly, \cite{GodinChapman2003} have shown that coupling improves in shallow waters due to resonant effects associated with marine sediments. These studies suggest that under the right conditions airwaves from fireballs might be detectable as hydroacoustic signals. 

However, few works have examined air-water shock coupling at the energies ($\sim$ kilotons of TNT = 4.185$\times 10^{12}$ J) and source altitudes (tens of kilometers) appropriate for bolides. \cite{Clarke1996} performed numerical simulations of shock coupling to water  from near surface nuclear explosions. That work showed that shock coupling decreased with increasing burst height. The result of this modeling suggests that a 1 kT airburst at 1 km altitude transmits the equivalent energy of a 10-50 kg TNT detonation in the SOFAR channel with peak energy between 2-10 Hz. This represents a net energy transmission to the SOFAR channel of order 10$^{-5}$ from a surface burst. 

We note that even if this energy coupling is several orders of magnitude lower at higher altitudes, the energy deposition would still be of order a kilogram TNT in the SOFAR channel from multi-kT airbursts. This is commonly cited as the limit for hydrophone detection at ranges of order 1000 km \citep{Wang2024}

From the foregoing, given the large theoretical and observational uncertainties, it appears feasible to search for the shock waves produced by fireballs, provided they are sufficiently energetic. We expect such hydroacoustic coupled explosive waves will be more detectable if the explosive source occurs over wavy ocean surface at shallow depths, at lower altitudes and closer to existing hydrophones.  As an approximately one kiloton or larger fireball occurs globally every 3-4 months \citep{Brown2002} there should be many opportunities for hydroacoustic detection of fireballs. 

While we have to this point focused on detecting fireballs via airborne explosion shocks coupling to the ocean, large meteorites surviving fireball ablation and impacting the ocean surface might also be expected to produce a hydroacoustic signal. \cite{Leslie1964} finds roughly 0.1\% - 1\% of initial kinetic energy is transferred from a projectile impact site at the ocean surface into hydroacoustic energy, but does not explore how this signal attenuates with depth. 

The purpose of this work is to perform a preliminary reconnaissance survey of fireballs occurring over the ocean which might be expected to produced hydroacoustic signals, search for those signals and try to establish a causal link. Our goal is to explore the expected characteristics needed for a fireball to be detected hydroacoustically and to search for such signals. 

In doing this survey we explore the indirect signals from bolides. We assume energy deposited by the fireball (either through coupled shock waves or meteorite impacts) is trapped and at some distance from the source is detected in the SOFAR channel. Underwater pressure signals which penetrate to the depth of SOFAR channel may be detectable over large distances with a low transmission loss (see Section \ref{SOFAR_section}). Thus we expect a priori if a fireball acoustic shock incident on the water surface (or a hydroacoustic signal is produced by a meteorite impacting the ocean surface) can penetrate to the depth of the SOFAR channel it is likely the signal could be detected by a hydrophone at a range of order several thousands of kilometers. Our analysis is limited to hydroacoustic signals propagating directly to hydrophone arrays which are part of the International Monitoring System. We also restrict our analysis to a default set of parameter choices in the signal processing software (DTK-GPMCC - see section \ref{methods} for details) for detection, which includes fixed time windowing, time overlap, discrete filter bands and a specific threshold in consistency per arrival family. Our results strictly apply only to signals processed in the same manner and hence the results should not be interpreted as anything but an order of magnitude detection limit.   

We present this analysis by starting in Section \ref{SOFAR_section} with a review of the basics of hydroacoustic propagation and introduce a simple pressure-yield scaling from earlier works. In Section \ref{source_mech} we explore the order of magnitude expected efficiency of hydroacoustic signal production from the shock wave produced by a fireball and from meteorites impacting the ocean surface. Section \ref{methods} details our methodology and initial survey to find potential fireballs to use in this study, and our method to determine if a hydroacoustic signal is fireball-related. Section \ref{results} and \ref{discussion} present and discuss the results of our survey.

\section{Hydroacoustic Detection and the SOFAR Channel} \label{SOFAR_section}

\begin{figure*}
  \includegraphics[width=\linewidth]{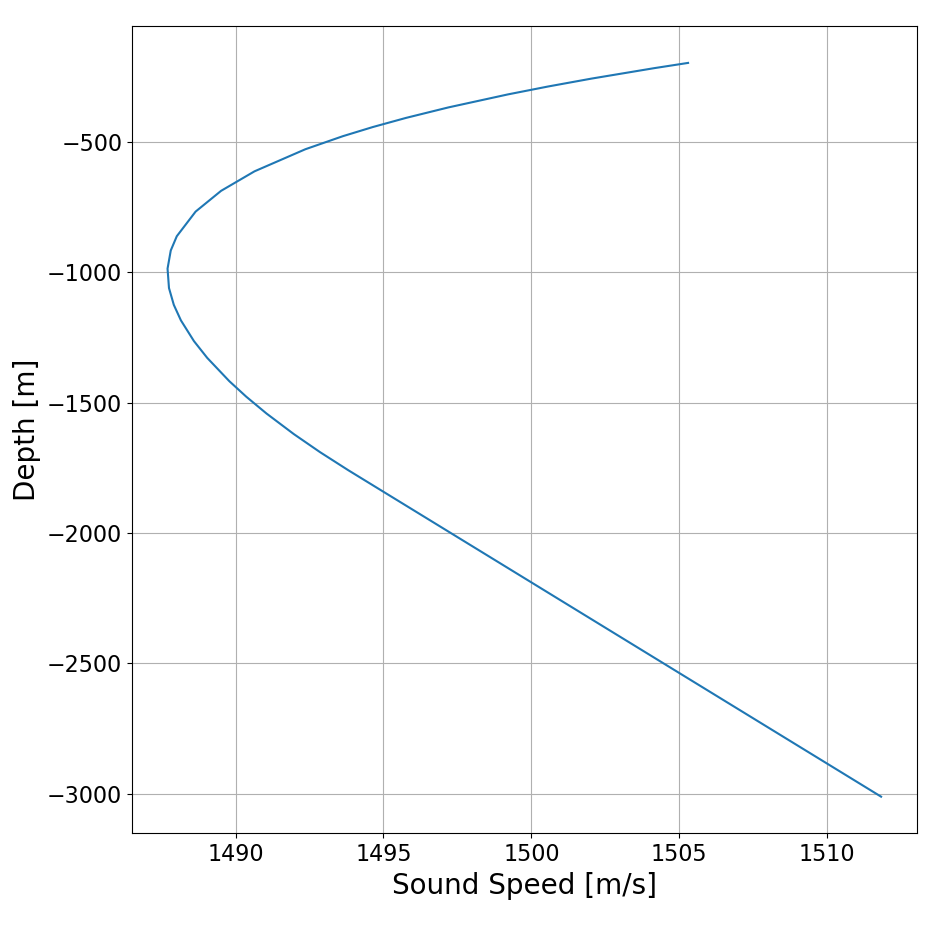}
  \caption{Figure adapted from a similar plot in \cite{Urick_1975}. This shows acoustic sound speed as a function of depth in the ocean. Note the minimum at around -900 m; this represents the sound fixing axis of the SOFAR channel.}
  \label{water_sound_profile}
\end{figure*}

Long range underwater signal propagation in the ocean is determined by the deep sound channel or Sound Fixing and Ranging (SOFAR) channel \citep{Jensen_2011}. This is a zone found at varying depth, depending on global location and time of year, where signals propagate through refraction within a waveguide. It is typically located 0.3 - 4 km below the ocean surface \citep{Northrop1975} with the zone approaching the surface near the poles. The depth at which the minimum sound speed occurs is termed the sound channel axis, and it is found at approximately 1 km in depth \citep{Urick_1975} at mid-latitudes. 

Figure \ref{water_sound_profile} shows a sample water sound profile. In analogy to the variation in atmospheric sound speed with height, which facilitates infrasound ducting near sound speed minima \citep{Lepichon2019}, the sound speed variation in the ocean allows for hydroacoustic signals to be trapped near the depth of sound speed minimum. This allows signal propagation over large distances (thousands of kilometers).
For reference, the transmission loss in the SOFAR channel is roughly:
\begin{equation}
    T = 10\log(r)
\end{equation}
where $T$ is the energy transmission loss (for cylindrical spreading) in deciBels (dB) relative to 1 m distance, and $r$ is distance measured in km \citep{Gerstoft1999}. From this relation, hydroacoustic signals traveling distances of 10000 km will be reduced in amplitude to 1\% of the original value at the source (-40 dB in energy). 

Hydrophone stations are expensive to deploy and operate; hence few exist. Global hydroacoustic data have become more available since the Comprehensive Nuclear-Test-Ban Treaty Organisation (CTBTO) was formed in 1996. As part of the CTBTO seismic, infrasound, hydroacoustic, and radionuclide stations have been deployed in the International Monitoring System (IMS) with the goal of global detection of any nuclear detonations. While there are more than 170 seismic stations used by the IMS (in addition to thousands more seismometers operated by regional and national networks) we confine our interest to the IMS distribution of active infrasound, hydroacoustic and T-phase stations the locations of which are shown in Figure \ref{ctbto_ims} \citep{Lepichon2019}). T-phase stations are seismic stations located near ocean interfaces where steep topography permits coupling between the ocean and the solid Earth allowing stronger hydroacoustic signals to be detected as seismic waves \citep{Okal_2001}.

\begin{figure*}
  \includegraphics[width=\linewidth]{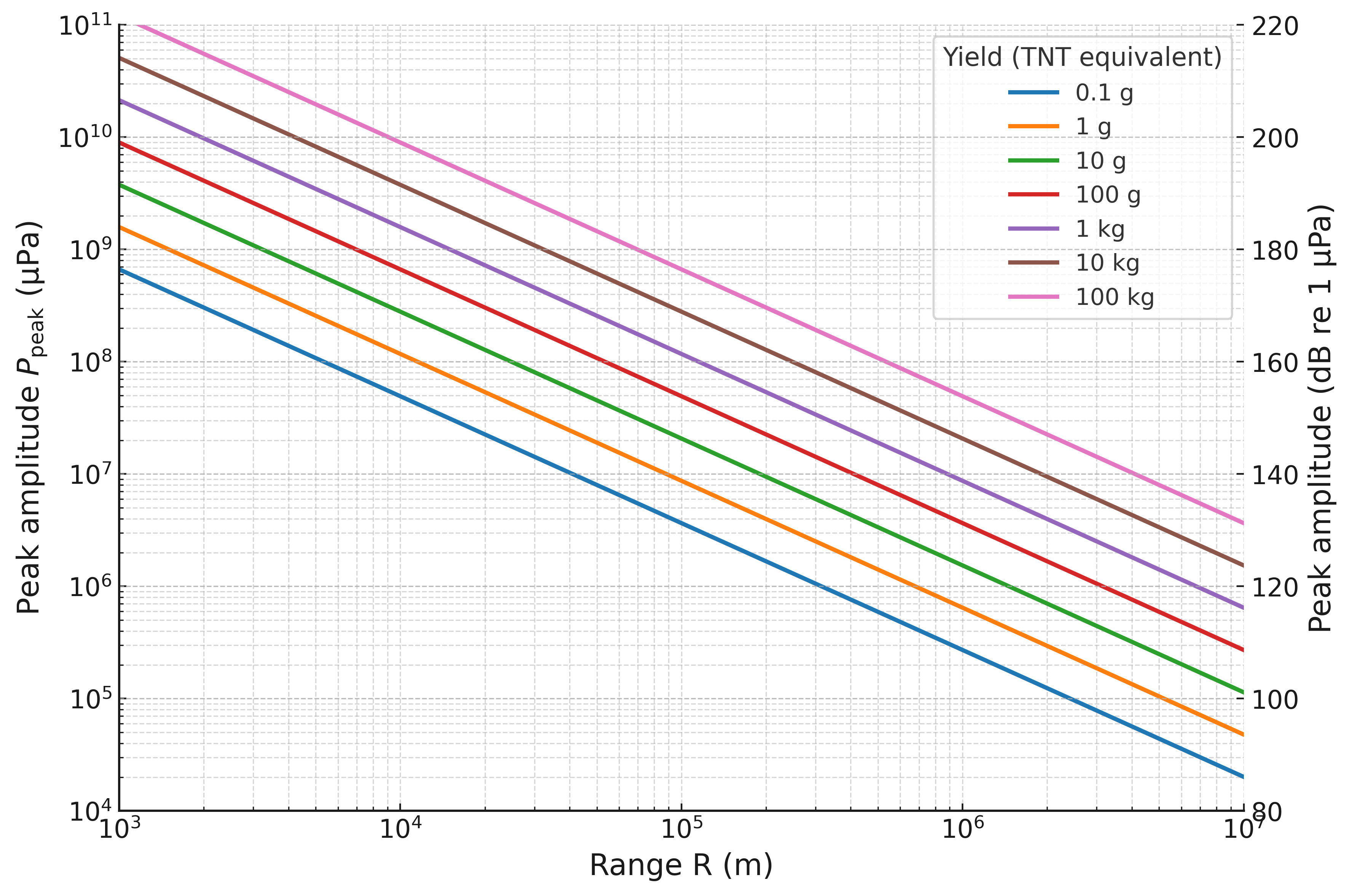}
  \caption{The estimated peak pressure as a function of range for different TNT equivalent explosive yields following \citet{Arons_1954}. The relation assumes both source and receiver are in the SOFAR channel.}
  \label{fig:sofar/amplitude}
\end{figure*}

Hydrophones located in the SOFAR channel are detection limited by local noise. An empirical relation \citep{Arons_1954, Soloway2014} relating range (R in meters), yield (W in kg of TNT equivalent) and peak  amplitude (P$_{peak}$ in Pa) for explosions in the SOFAR channel given by

\begin{equation}
    P_{peak} = 52.4 \times 10^6(\frac{R}{W^{1/3}})^{-1.13}
\end{equation}
\label{eq:energy}

provides a convenient means of estimating the expected signal pressure for a given amount of energy deposited in the SOFAR channel as a function of distance. For a bandpass from 5-15 Hz, typical hydrophone pressure noise levels vary from 100 - 120 dB, though surface conditions and anthropogenic sources can drive levels an order of magnitude above or below this typical range \citep{Jensen_2011}. Based on Figure \ref{fig:sofar/amplitude} for characteristic source to receiver distances of order 1000 km, this suggests that source energies in the SOFAR channel need to be in the tens to hundreds of grams TNT range to be sufficiently above the noise floor to be detected.

In \ref{shot_cals} we search for literature examples of explosions with known location, timing and yield. Unfortunately, there are no modern published examples of SOFAR explosions with these quantities reported. As a result, we cannot fully validate the use of Eq. \ref{eq:energy}, but note that the results from \ref{shot_cals} show that for explosions at shallow depth outside the SOFAR channel this relation typically over predicts the expected pressure at long ranges. Thus our use of this relation should be considered an upper limit to true yield for explosions occurring outside the SOFAR channel. 

In our work, we assume the SOFAR channel is the primary method of hydroacoustic transmission of a signal originating from a fireball to a hydrophone array. We characterize the depth of the SOFAR channel for a specific latitude and longitude as the depth at which the minimum hydroacoustic sound speed occurs, based on the model of \citet{Chu_Fan_2024} which uses the temperature and salinity profile in the ocean.

\section{Fireball Hydroacoustic Source Mechanisms} \label{source_mech}

Detecting a fireball signal using hydrophones located in the SOFAR channel requires that some fraction of the fireball energy deposited in the atmosphere be transmitted across the air-ocean inferface and into the SOFAR channel. 

This energy transmission can occur through two possible pathways. The first is through direct coupling of the shock wave generated by the fireball when the shock reaches the air-water interface. The second mode is direct production of hydroacoustic signals when any meteorites which survive the fireball physically impact the ocean surface moving at high speed. We will review each of these possibilities in turn. 

\subsection{Shock wave transmission through the Air-Water Interface}
For an acoustic wave in air incident on the air-water interface, the vast majority of the energy is reflected, due to the large impedance differences between the two media \citep{Benmenahem1981}. Based on measurements and theoretical considerations of the sonic boom created by aircraft over the ocean surface, we expect a small portion, roughly 0.1\% of the incident energy of the shock, to be transmitted directly through the air-water interface \citep{Sawyers1968, Cook_1970, Chapman1990}, though this can be up to an order of magnitude higher if the sea surface is rough \citep{Meecham_1976}. The signal will suffer further attenuation with higher frequencies preferentially removed as it propagates to the depth of the SOFAR channel \citep{Sohn2000, DesharnaisChapman2002}.

However, due to the low-attenuation propagation loss once in the SOFAR channel, if a surface shock can deposit even a small amount of energy into the channel, the disturbance should be detectable over long distances \citep{Urick_1975}. 
Prior work related to horizontally flying aircraft has shown that objects moving at higher Mach number will have a relatively higher fraction of energy transmitted than reflected \citep{Cook1970} Note, however, this applies to the comparatively near field N-wave shocks which are of much higher frequency than is typically produced from large meteors, which may have dominant periods in excess of 5 seconds for kiloton-class events \citep{Revelle1976}.

\citet{Sparrow_1995} shows explicitly that the penetration depth and equivalent pressure for a shock incident at the ocean surface increases as the dominant frequency decreases. As typical sonic boom shocks have fundamental periods of order a few tenths of a second \citep{Carlson_1978} and are detectable above ambient ocean noise to depths of less than 100m \citep{Baird2025}, the larger periods of bolides qualitatively suggest penetration to several hundred meters is plausible.  \par

\subsubsection{Meteor Shock Wave Energetics}
\label{sec:shock_energetics}
To provide order of magnitude estimates of the energy impinging on the ocean surface from a bolide shockwave we will adopt as a standard fireball ``energy" a deposition into the atmosphere of 1 kT TNT ($4.184 \times 10^{12}$ J). This corresponds to roughly a 1~m radius object impacting the upper atmosphere at a typical speed of 20 km/s. Such objects impact Earth approximately once every three months \citep{Brown2002}. 
They also have characteristic heights of peak brightness centered around 25-35 km; we will adopt the lower bound of this range to set an upper limit for our standard fireball shock overpressure \citep{Brown2016, Silber2025}. 

To estimate the overpressure of the shock, we will further assume for simplicity that all of the fireball energy is released at the height of peak brightness - ie. we treat the fireball as a spherical point explosion an approximation commonly made in planetary defence modeling of energetic fireballs \citep{Aftosmis_2019, Wheeler_2024}. As shown by \citet{Collins2017, Shuvalov2013}, the point source approximation (adopted from the nuclear explosion literature \cite{Glasstone1977}) typically underestimates the ground overpressure compared to real bolides which have energy deposited over a cylindrical length of trail by a factor of two. Hence our adopted overpressure values are lower bounds for our chosen height. The resulting coupling coefficient will similarly be upper bounds. 

We follow the general methodology of \cite{Mcfadden2021} and use blast scaling from \cite{Kinney1985} for a spherical atmospheric explosion of 1 kT TNT at 25 km altitude to find an estimate of the overpressure at the ground directly beneath the explosion of order 100 Pa. We note that the more detailed model fit of nuclear test data given by \citet{Glasstone1977} across a range of heights by \cite{Collins2017} predicts a peak overpressure of 200 Pa for a point source explosion of this yield and at this height.  This is comparable to the ground overpressures estimated by others of the shock waves produced by specific kT-class fireballs. For example,  \citet{Gi2018} analysed actual observed fireball energy deposition profiles for bolides in this energy range and together with a weak shock numerical model to arrive at similar overpressure values. 

To compare with the energy of a discrete meteoroid impact, the energy density of the shock wave, $\frac{E}{V}$, can be written in terms of shock front area, $A$. We assume the thickness of this shock (the distance between the shock front and the rarefaction phase which occurs a short time later) is some number, $n$, of wavelengths, $\lambda$, long. Here the wavelength represents the fundamental (dominant) period of the shock. This normalization is useful to estimate a total energy to compare with the meteorite impact calculation in the next section. We could use intensity here, but since we approximate the energy released by the fireball as being from a single fragmentation episode, we remove the temporal dependence to arrive at:

\begin{equation}
    \frac{E}{A (n \lambda)} = \frac{P^2}{Z_{air} c_{air}}
\end{equation}
\label{eq:energy_shock}

\begin{equation}
    \frac{E}{A} = \frac{P^2 n}{Z_{air} \nu}
\end{equation}
\label{eq:energy_shock2}

where, $\frac{E}{A}$ is the wave energy per unit area, $P$ is the overpressure at ground level, $\nu$ is the frequency of the sound, Z$_{air}$ is the specific impedance of air and $c_{air}$ is the speed of sound. 

This represents the energy per unit area in the shock just before it reaches the ocean surface. The energy transmission coefficient between air and water for an acoustic wave, $T$, can be found from the standard wave transmission equation for energy as \citep{hecht2017}:
\begin{equation}
    T = \frac{4 Z_{air}Z_{water}}{\left(Z_{water} + Z_{air}\right)^2}.
\end{equation}
From which we get an energy transmission coefficient of $1.12 \times 10^{-3}$, using a specific impedance of 1.48 MPa s/m for water, and 413 Pa s/m for air. Note this is a very crude zeroth order approximation - we ignore the angular dependence on transmission, the surface roughness of the ocean and treat the wave as vertically incident. This results in an energy per unit area just below the ocean surface of:
\begin{equation}
\frac{E}{A} = T \frac{P^2 n}{Z_{air} \nu}.
\end{equation}
Solving this, we get $\frac{E}{A} \approx 10^{-3} \frac{n}{\nu}$. For a 1 kT explosion the dominant period is roughly 5 sec in the atmosphere \citep{ReVelle1997}, so $\nu$ is about $0.2$ Hz. Taking $n$ to be $1$ appropriate to a minimum energy density estimate, we get an acoustic wave energy per unit area just below the ocean surface on the order of 0.1 J/m$^2$ transmitted by the fireball shock wave directly beneath the point of detonation.

For comparison, \cite{Kamegai1994} performed numerical calculations of shock coupling to the water surface from airborne nuclear detonations in the deep ocean. They found that the coupling efficiency for low altitude bursts (40m above the ocean surface) was of order $10^{-4}$ which decreased further with increasing burst height. \citet{Sakurai_1969} developed an analytic theory of shock propagation between air and water and finds an overall energy transmittance of 10$^{-3}$. Thus our simple transmission assumption is order of magnitude comparable to previous numerical/analytic estimates for shock waves at the air-ocean interface. Note this does not account for further attenuation of the wave as it propagates downward to the SOFAR channel, so is an upper limit for our purposes. 

The total energy for the entire event transmitted is then a function of the shock footprint area. For our nominal 1 kT airburst at 25 km, the overpressure drops by roughly a factor of two over an area of order 10$^4$ km$^2$ about the fireball sub-terminal point. Thus the total energy imparted by the shock to the ocean is of order 10$^9$ J spread over approximately 1.5 minutes for an average source power near 10$^7$ W. 

Note that the foregoing simplified analysis neglects the effects of the changing incidence angle with fireball footprint geometry, ocean surface roughness, different dominant periods, varying burst altitudes, ocean depth as well as specifics of the bathymetry and transmission loss along the course to receiver path in the SOFAR channel. All of the above will modify the coupling and our quoted $\eta$ values should be interpreted solely as the relative efficiency of bolide signals to be detectable using our chosen PMCC parameters, simplified assumptions concerning height and energy release and the implicit modal assumptions incorporated in the use of Eq \ref{eq:energy}. 

\subsection{Meteorite Impacting the Ocean Surface}
High velocity objects, such as missiles or re-entry vehicles, impacting the ocean can produce locally detectable hydroacoustic signals \citep{stumpf2021}. For fireballs, in many cases, there is significant material which survives the ablation process after the luminous phase of flight. This debris, in the form of meteorites, will impact the ocean surface moving at terminal velocities  of order tens to a few hundred m/s for kilogram to hundreds of kilogram fragments \citep{Ceplecha1998}. Therefore it is worth examining if it is possible for hydroacoustic waves generated in the water through a meteorite impact to be sufficiently energetic to trap significant energy in the SOFAR channel and be detected at long ranges. 

One possible approach to distinguish shock wave vs meteorite impacts as sources could be time delays.  For the case of a fireball impacting vertically and then fragmenting at 25 km altitude, large fragments (of order 100 kg) will decelerate to speeds below 3 km/s in just a few seconds. These will further decelerate and fall at terminal speeds taking approximately one minute to reach the ocean surface. Smaller masses will take considerably longer (up to 3 minutes for kilogram sizes and 10 minutes for gram-sized masses) but these will impact at much lower speeds (a few tens of m/s compared to more than 100 m/s for hundred kilo masses) and contribute negligibly to any hydroacoustic signal. 
 
 For comparison, the shock wave originating from this height will take roughly one minute to reach the water surface. For typical source-receiver separations both signals will travel hydroacoustically for of order 30 minutes and thus it would be nearly impossible to resolve the meteorite impact and the shock wave source through temporal arguments alone. 
 
 As a reminder, in this work, we represent the bolide as a point energy source occurring at its height of peak brightness, consistent with the notion that many energetic fireballs deposit energy in a short portion of their trajectory once fragmentation commences (an airburst) \citep{Wheeler2018}. We thus only consider the shock wave time to reach the water surface. 

If the same meteoroid we considered in section \ref{sec:shock_energetics} were to impact the ocean's surface, we can estimate the energy deposited to the water. For this calculation we need to estimate the mass of meteorites which survive ablation to impact the water.  Note that with the possible exception of rare iron meteoroids we do not expect meter-sized bodies to reach the ocean surface moving at supersonic velocities but rather to fragment and deposit meteorites which impact at terminal speeds. 

For a typical stony meteoroid, a 1 kT fireball corresponds to an initial mass of  approximately $2 \times 10^4$ kg, assuming a speed of 20 km/s. Following the ablation model and survival fraction developed by \cite{Bland2006}, this would deposit on the surface numerous stone fragments over a strewnfield of order a few tens to hundreds of km$^2$ filled with an average fall masses of ~1-10 kilograms per km$^2$. This corresponds to an energy density of approximately $10^{-2}$ to $10^{-3}$ J/m$^2$, distributed mainly in a few large kilo or tens of kilo masses. These will be traveling on the order of 100 m/s upon impacting the ocean surface and couple each generate a potential hydroacoustic signal. 

\begin{figure*}
  \includegraphics[width=\linewidth]{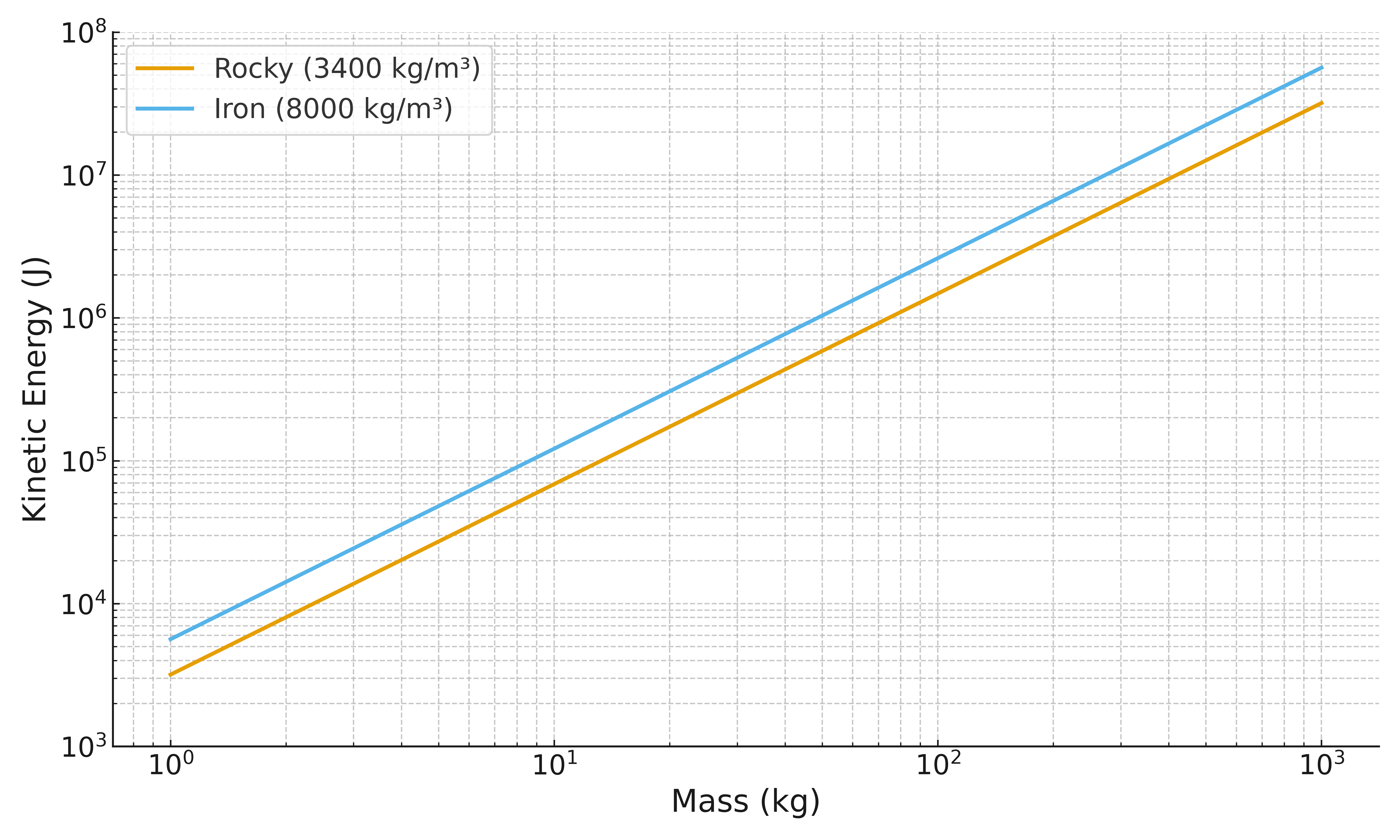}
  \caption{Terminal impact energy of different sized spherical iron and stony meteorites.}
  \label{fig:terminal-speed}
\end{figure*}

Figure \ref{fig:terminal-speed} shows the expected kinetic energy for terminal velocity fragments from stony and iron meteorite fragments. Note that the largest individual stony meteorite ever recovered as a single mass was the Jilin, China fall of 1976 where a single 1770 kg was recovered. Thus the upper limit to single stones is of order 1000 kg, though such large singular masses are rare. A typical kT - class stony impactor  has as its single largest fragment a meteorite with mass about 0.1 - 1\% of its initial mass \citep{Bland2006} corresponding to a fragment mass of 20 - 200 kg. Such a  typical single fragment will impact with of order 10$^6$ J of kinetic energy. We focus on the effect of the single largest fragment, recognizing the stochastic nature of fireball fragmentation to set a limit in terms of the largest individual impulsive event which could reasonably be expected from a fireball. 

By comparison, the energy deposited by an iron impactor to the surface is about three orders of magnitude larger than a similar energy stony impactor \citep{Bland_Artemieva_2003}. For a 1 kT initial energy iron impactor, we expect the largest fragment to be of order $10^3$ kg \citep{Bland2006} and impact with a velocity of about 1 km/s. This corresponds to an energy of 10$^9$J, equivalent to the entire shock energy coupled to the ocean for our standard 1 kT fireball, but as a singular event and hence with much higher power. However, as iron meteorite falls represent only about 4\% of all meteorite falls \footnote{https://www.lpi.usra.edu/meteor/metbull.cfm} and assuming this fraction can be extrapolated to meter-sized impacting bodies this would suggest iron impacts in the kT-class are rare, occurring globally only once every 5 - 10 years \citep{Brown2002}.

From the foregoing analyses, the energy density that penetrates the water from a bolide shock wave is typically many orders of magnitude larger than for all the energy deposited from a single large stony meteorite impact. Hence for most fireballs we expect the shock wave to be the most likely to inject energy into the SOFAR channel to be detected at long range. In the case of an iron impactor, the energy reaching the ground is much larger than a stony and may be comparable to that expected from shock wave coupling, but these are very rare. With these energies in mind, \cite{Leslie1964} finds roughly 0.1\% - 1\% of initial kinetic energy is transferred from a projectile impact site at the ocean surface into hydroacoustic energy, suggesting that for large meteorites sufficient energy could be trapped in the SOFAR channel to be detectable at long ranges.

The actual efficacy of hydroacoustic signal production by a meteorite impacting the ocean is best constrained through a similar real-world example. There are at least 10 examples of long range hydroacoustic signals produced from surface airplane impacts \citep{Kadri_2024}, many of which show signals consistent with lab measurements of impacts onto water surface \citep{Kadri2017}. These have comparable energies to meteorite impacts, though specific information on impacting energy is lacking for most cases. 

Among these aero-ocean impacts, the best constrained is an F-35 high speed crash off the coast of Japan on April 9, 2019. This was recorded at one IMS hydroacoustic station and confirmed at a set of hydrophones near the Japanese coast \citep{Metz2022}. Among all airplane ocean impact events summarized by \citet{Kadri_2024} this has the highest signal to noise, is impulsive, and is most convincing in that multiple sites detected the event in the right time window from the correct backazimuth with consistent signal features.  

Most significantly, the accident report notes that the impact speed was known and together with the F-35 mass (and bounding mass values of fuel remaining at the time of the accident) provides an estimate for the impacting kinetic energy of 900 $\pm$ 200 MJ. This is comparable to the energy of an iron meteorite impactor \citep{Bland_Artemieva_2003} but is 2-3 orders of magnitude above the energy expected for a typical large stony fragment from a 1 kT bolide, given that only half a dozen stony meteorite falls have ever produced individual fragments exceeding 100 kg\footnote{https://www.lpi.usra.edu/meteor/}. 

This impact is the closest real-world analog to a high speed meteorite impact into the ocean with published information on the energy and location. We note, however, that a high speed aircraft oceanic impact would have different coupling to a rocky impact because of differences due to potential aircraft breakup, fuel, impact angle and the penetration depth-time history of the airframe.  It is not a good analog to what would be expected for air shock coupling and so the comparison should remain strictly with meteorite ocean surface impacts. 

For the F-35 crash, using this impact energy, the observed peak pressure of 0.7 Pa at H11 (one of the CTBTO hydroacoustic stations located at Wake Island), the known range of 3300 km and Equation \ref{eq:energy} the resulting coupling efficiency from the surface to the SOFAR channel is of order 10$^{-4}$. Note, however, that at the crash location near the Japanese coast the SOFAR channel is estimated to only have a depth of 500-700m \citep{Chu_Fan_2024}, so this may be a higher efficiency than in the open ocean and again should be taken as an upper bound.

Using the analysis from \citet{Metz2022}, the impulsive signal at H11 is one order of magnitude in pressure amplitude above the local background noise at the time of the event. Using the energy-pressure-range relation in Figure \ref{fig:sofar/amplitude} we find that an impulsive impact with energy of 1 MJ with identical coupling properties into the SOFAR channel  would be near the noise floor for this example. Thus the typical largest stony meteorite fragment from a 1 kT bolide impacting the ocean in this order of magnitude scenario would be close to the noise level and likely not detected, particularly given that the coupling efficiency is an upper limit.  

We conclude that for the majority of fireballs the most probable mechanism for hydroacoustic sound detection of a fireball is from the airwave shock coupling to the water surface. The exception being rare iron impactors where individual fragments would impart significant energy upon impacting the ocean surface.

\section{Survey Methodology}\label{methods}

To search for potential candidate fireballs producing hydroacoustic signals, we assume that the most probable detections would be from one of three potential fireball populations. These populations include the most energetic bolides (E $>$ 5 kT) appearing over open ocean, fireballs impacting in ocean regions where the SOFAR channel comes close ($<$ 500 m) to the ocean surface and fireballs occurring close ($<$500 km range) to a hydroacoustic station (HS). 

Uncertainties in timing and location for CNEOS events have been estimated through comparison to common ground-based fireball detections. For timing uncertainty, \citet{Peña-Asensio_2025} found a maximum difference between reported sources of less than 12 seconds, while \citet{Devillepoix2019} found absolute timing agreement to be of order a few seconds or better in most cases. The expected absolute origin time uncertainty for CNEOS fireballs of order a few seconds, is much smaller than the uncertainty in propagation time for the shock from the fireball to the ocean surface where we have assumed a mean acoustic speed of 0.31 km/s across all heights, while the true average signal speed will vary with source height and temperature profile, such that mean speeds over height ranges of interest for bolides in the energy class we examine may vary from 0.34 - 0.28 km/s depending on height \citep{Waxler2019}. This results in a typical uncertainty in time from shock production to first reaching the ocean surface of order 10 seconds. The contribution to the timing uncertainty in our survey is thus dominated by our large range in hydroacoustic celerities which produce acceptance windows of order many minutes for most long range fireballs (see Tables \ref{tab:cneos_high_energy} and \ref{tab:cneos_shallow_SOFAR}).

To compile a list of candidate fireballs in these three categories during the $\approx$25 years of operation of the IMS hydroacoustic network, we examined all fireballs listed on the Center for Near Earth Object (CNEOS)\footnote{https://cneos.jpl.nasa.gov/fireballs/} website. This source provides information on bright fireballs (roughly 1 meter or larger in diameter) detected globally by US Government sensors \citep{Brown2016}. Data for each bolide include information on location, timing, height of peak brightness, total energy as well as velocity, but without uncertainties \citep{Peña-Asensio_2025}. 

We further restricted our dataset to the CNEOS fireballs that occur over water using the Pyproj\footnote{https://pyproj4.github.io/pyproj/stable/} library, by comparing the geographic coordinates given by CNEOS to the water polygons available in Pyproj. We then categorzied all events as being in one of our three populations (energetic, shallow SOFAR or proximal to hydroacoustic stations). Note that we do not use T-phase stations but only the six IMS hydrophone arrays as our survey is restricted to direct H-phases only.  Starting with this list we further reduced our event search to include only those fireballs which had an approximate great circle water path uninterrupted by land from the source position to at least one of the IMS hydroacoustic stations (see Figure \ref{ctbto_ims}).

For our candidate fireballs, we then compute the expected arrival time and azimuth for any fireball signal using the geographical distance from the fireball to each HS and a fixed speed of sound in water of 1.48 km/s appropriate to the SOFAR channel \citep{Urick_1975}. We include in this delay time the time for the shock wave to travel from the height of the brightest point on the fireball to the ground and reach the water surface assuming a fixed sound speed of 0.31 km/s. This produced an expected approximate time of signal arrival at each HS. To allow for possible variable hydroacoustic propagation speeds, we computed an acceptance window using the distance from the fireball to the HS and a celerity ranging between 1.42 - 1.55 km/s. This celerity range incorporates the majority of observed signal speeds for H-phases across latitudes and seasons reported in the literature (eg. \citep{Evers2015, Mikhalevsky2001}), while azimuthal angular deviations for H-phases for explosive sources are less than a degree at long ranges \citep{Prior2010} in agreement with the azimuthal deviations we find in \ref{shot_cals} which average less than 0.3 degrees.

Using the expected arrival time as the center of a 1h window, we examined IMS hydroacoustic data from each HS station. This time window was many times larger than the largest acceptance window. It was chosen to allow for an estimate of the background signal rate near the expected arrival time at each HS. 

We found most 1h windows of hydroacoustic data had multiple signals. Many were from very similar azimuths. We therefore removed these repetitive signals from further consideration as being associated with a fireball. A signal was classified as repetitive and removed if three or more arrivals occurred in the 1h window coming from within one degree of the same azimuth. 

A fireball signal was provisionally associated with a hydroacoustic signal if it was within the acceptance time window and was also within $\pm$ 5$^{\circ}$ of the expected great circle backazimuth.

Potential signals were identified using the PMCC (Progressive Multi-Channel Correlation) algorithm \citep{Cansi1995, Cansi2008} to analyze the waveforms. The Dase Tool Kit-Graphical Progressive Multi-Channel Correlation (DTK-GPMCC) software program (v6.11.5) was applied to all HS data. We chose 10 linearly spaced filter bands between 1 - 40 Hz with window lengths of 15 seconds and 90\% overlap. Pixel detections required consistency of 0.05. Groups of pixels were associated as arrival families using the default family association settings in v6.11.5 of DTK-GPMCC. These parameter choices were chosen following previous forensic work on hydroacoustic detections associated with the ARA San Juan submarine \citep{Vergoz2021, Nielsen2021}.

The resulting output consists of linked pixels in time - frequency space (families) where all three array elements show correlation above noise. Under the assumption of plane-wave propagation, the time delay arrivals across the 3 element arrays is then used to compute a signal backazimuth and trace velocity for each pixel. Each pixel value of these quantities is then used to produce family averages for these metrics. It is these family-level averaged signal values we use in our survey.

\section{Results}\label{results}

Tables \ref{tab:cneos_high_energy}, \ref{tab:cneos_near_stations} and \ref{tab:cneos_shallow_SOFAR} summarizes all 30 fireballs meeting our inclusion criteria and were either high energy (E$>$5 kT) [11 events], near hydroacoustic stations ($\leq$500 km range) [7 events] or occurred in oceanic regions where the SOFAR channel was shallow [12 events] respectively. The latter used the SOFAR channel depth model of \citet{Chu_Fan_2024} to identify where the SOFAR was within $\approx$ 500m of the ocean surface.

In total we found 15 PMCC families which occurred within the time acceptance window across all thirty fireballs (from among 53 potential HS stations where detections would be expected). However, of these 15, only one also had a backazimuth within $\pm$ 5$^{\circ}$ of that expected from the fireball in question. 

This signal was assosciated with a fireball which occurred on Sep 2, 2003, had an accepted signal on H03N (Juan Fernandez Archipelago) at a range of 12300 km. The associated fireball had an energy of 1 kT according to CNEOS. Infrasound associated with the fireball was detected at the IS10 infrasound array in Canada. Using the bolide infrasound analysis methodology summarized in \citet{Ens2012, Gi_Brown_2017} we find an average infrasound period at maximum signal amplitude of 4.3 $\pm$ 1.2 secs. Using the single station infrasound period-yield relation given in \citet{Ens2012} this corresponds to an energy of 0.8 $\pm$ 0.5 kT, consistent with the CNEOS energy estimate.

\begin{figure*}
  \includegraphics[width=\linewidth]{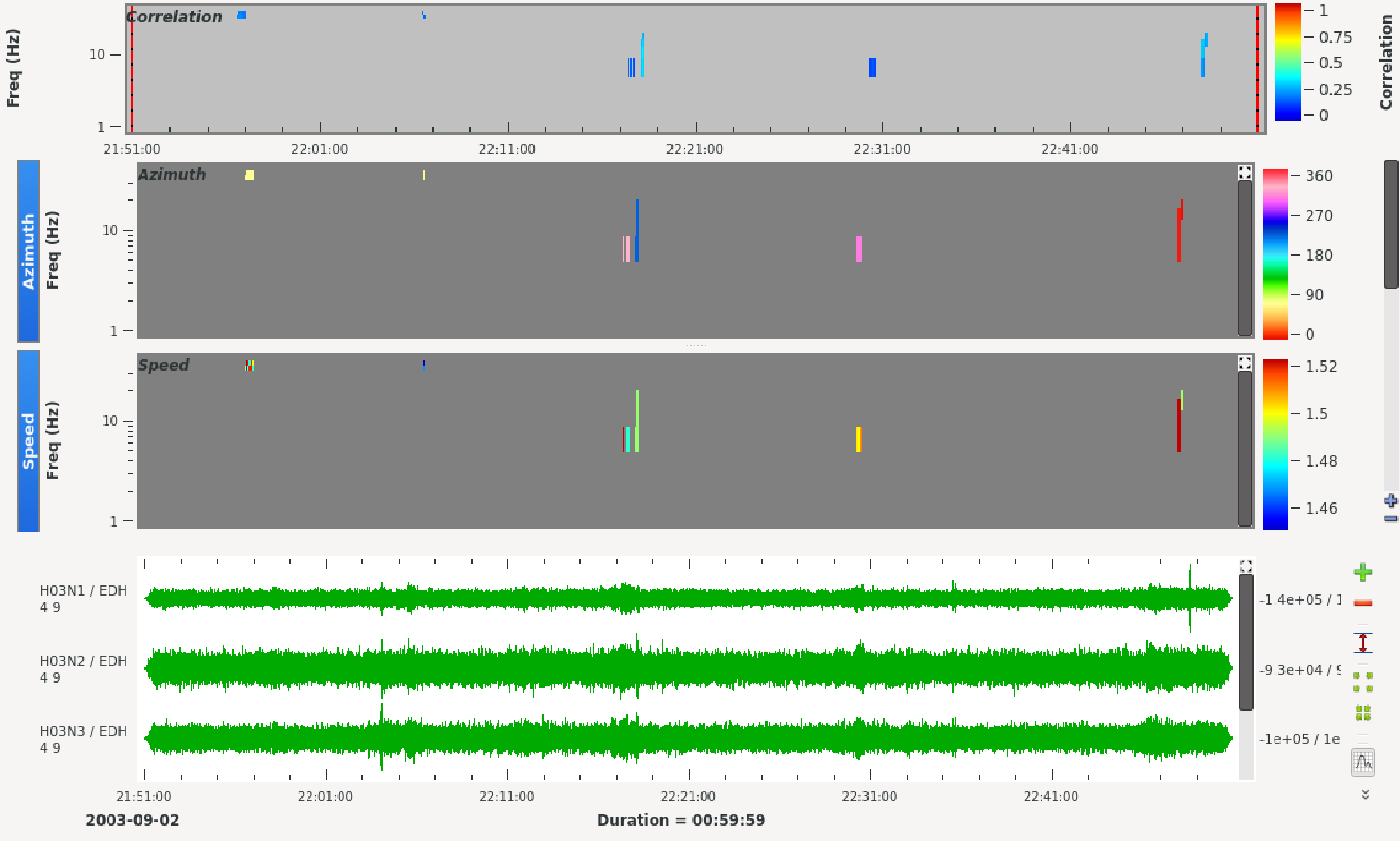}
  \caption{The PMCC 1h window from the H03N station for the September 2, 2003 event.  The top plot shows time-frequency blocks of family correlations, while the middle plots shows family arrival azimuths (color bar) for signals with high correlation. The lower plot shows the apparent trace speed (color bar in km/s) across the hydrophone array for these correlated signal "families".The associated signal occurs near 22:17 UT. The bottom three plots are the individual array element pressure-time signals.}
  \label{fig:20030902_bigwindow}
\end{figure*}

\begin{figure*}
  \includegraphics[width=\linewidth]{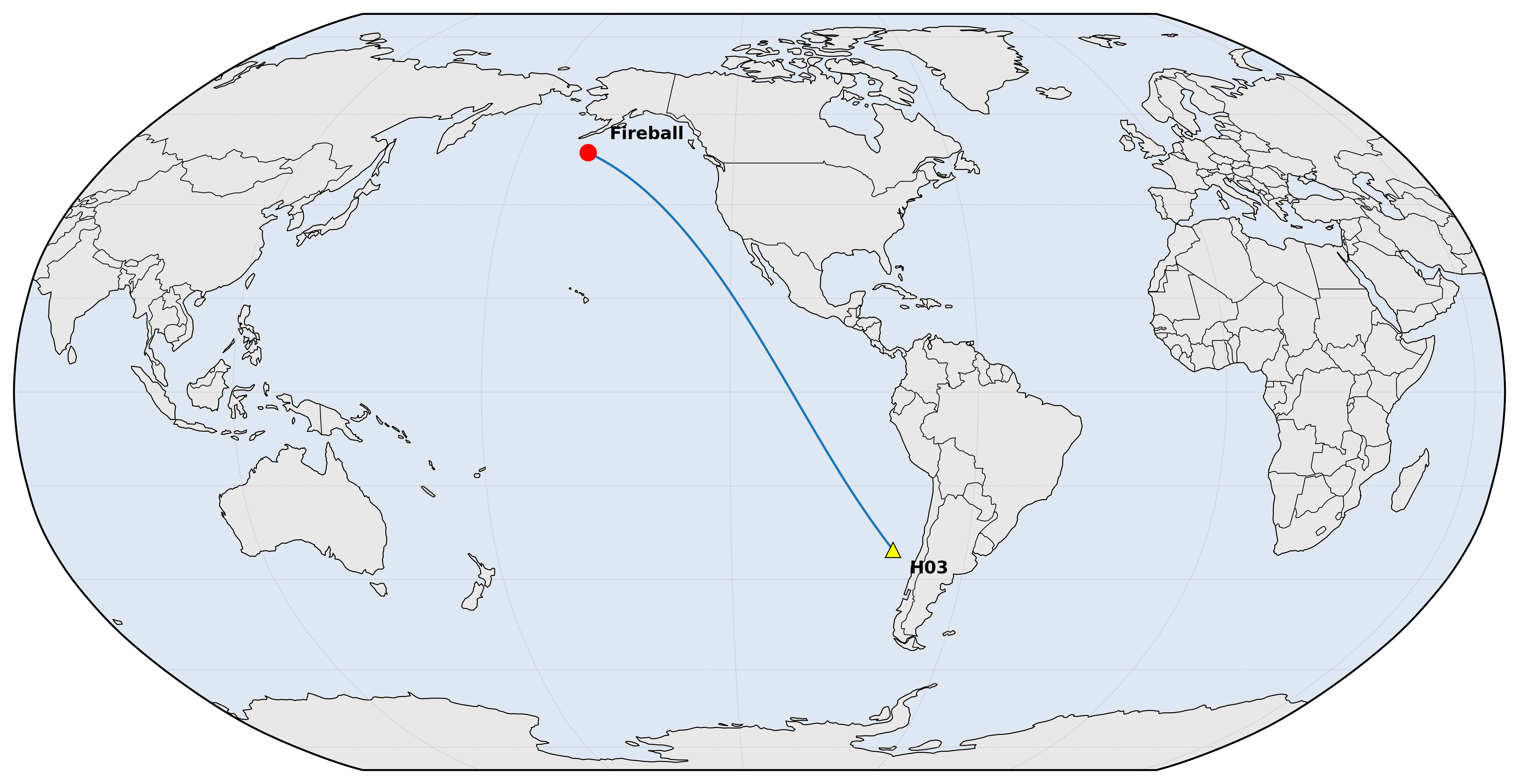}
  \caption{Location of the September 2, 2003 fireball in relation to HS station H03. The blue line shows the great circle path between the fireball and H03.}
  \label{fig:20030902_map}
\end{figure*}

The signal and associated waveform are shown in Figure \ref{fig:20030902_bigwindow} while the location of the fireball relative to H03N is shown in Figure \ref{fig:20030902_map}.

Within the acceptance window there are two signals. The possible fireball associated signal is expected to show a backazimuth of 318.4$^{\circ}$. One arrival occurs in the acceptance window at 22:18:05 UT but from an azimuth of 222.8 $^{\circ}$ well outside the expected azimuth range. 

The provisionally associated arrival at 22:17:24 shows an azimuth of 321.7$^{\circ}$ (3.3 $^{\circ}$ from the expected azimuth) and lasts for 26 sec. This brief, impulsive signal has a centre frequency near 7 Hz and noticeable energy between 4 - 8 Hz. The peak pressure is 0.04 Pa (or 40 000 $\mu$Pa), corresponding to a signal-to-noise ratio of less than 2.

Unfortunately, the only other HS array which had a possible direct path (H11 - Wake Island) was not recording data on this date. We are left with only a single hydrophone station detection of the potential signal.

\begin{figure*}
  \includegraphics[width=\linewidth]{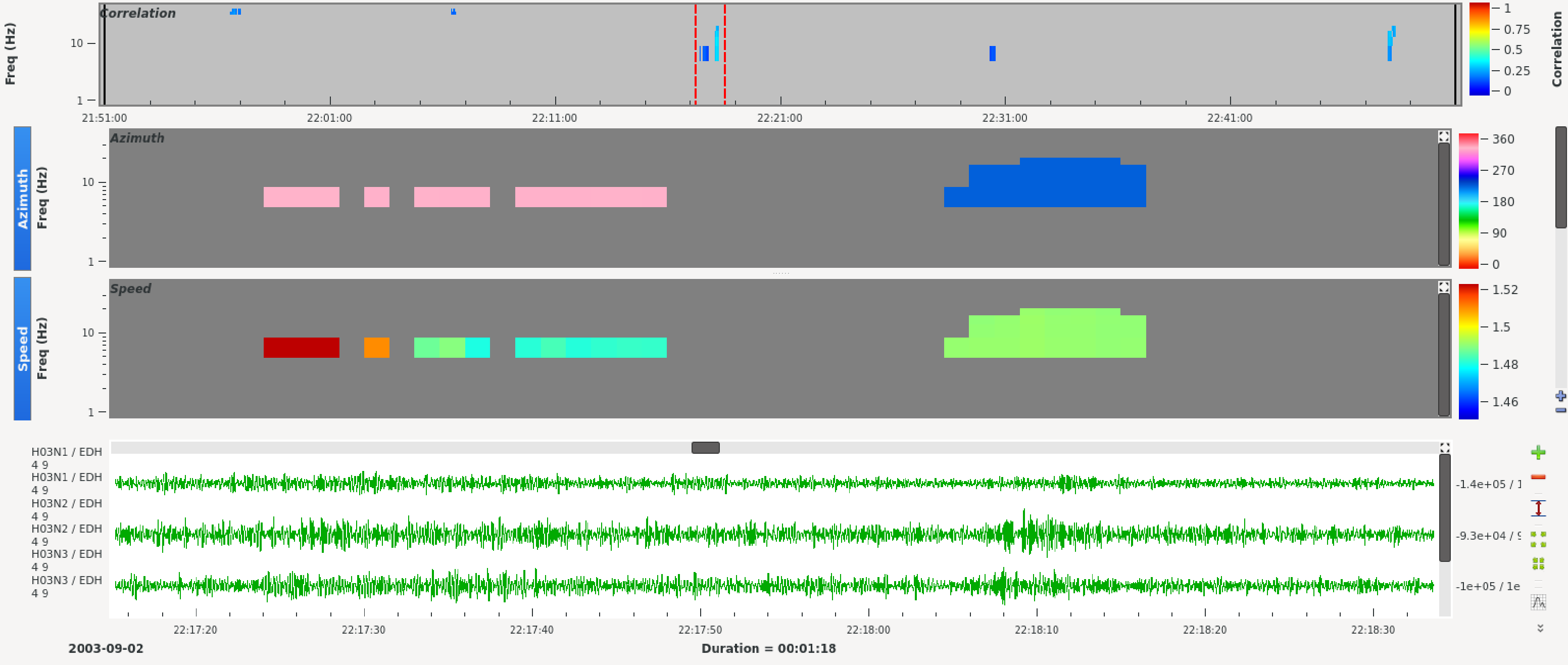}
  \caption{Same as figure \ref{fig:20030902_bigwindow} but zoomed in around the time of the signal and filtered from 4 - 9 Hz.}
  \label{fig:20030902_smallwindow}
\end{figure*}

\begin{figure*}
  \includegraphics[width=\linewidth]{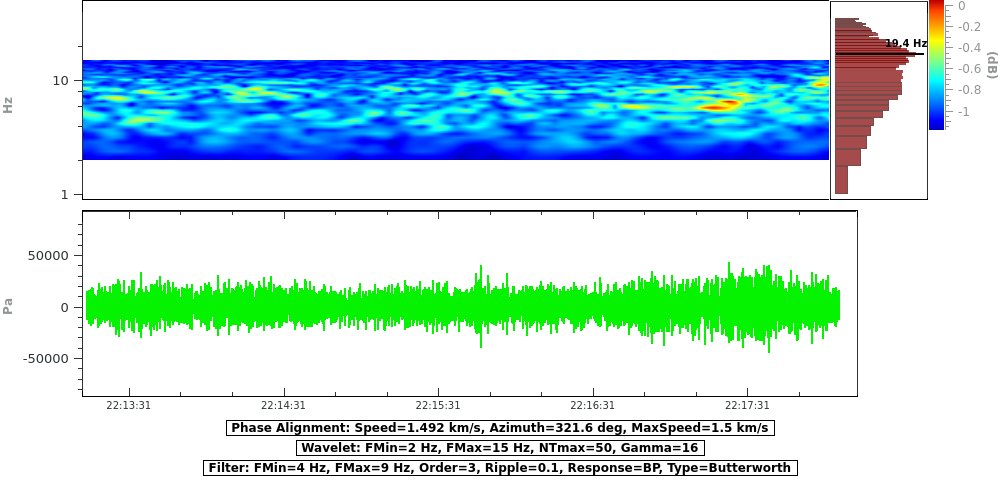}
  \caption{Stacked best beam (lower plot) filtered from 4 - 9 Hz for the signal provisionally associated with the September 2, 2003 fireball, which manifests as the slight increase in amplitude of the stacked signal peaking near 22:17:35. Also shown is the spectrogram for the signal.}
  \label{fig:20030902_spectra}
\end{figure*}

\begin{landscape}
\setlength{\tabcolsep}{4pt} 
\small  
\begin{adjustwidth}{-2.5cm}{0.5cm}
\begin{table}[h!]
\centering
{\renewcommand{\arraystretch}{0.6}
\begin{tabular}{llrrrrlrrlr}

\toprule
               Date &     Time (UT) &   Lat &   Long &  Ht (km) &  E (kT) & Station &  R(km) &  Az & Arr. Time &  Dur \\
\midrule
         2018-12-18 & 23:48:20 &  56.9 &  172.4 &         26.0 &         49.0 &    H11N &      4208.4 &          5.2 &             00:37:07 &                   4.14 \\
         2018-12-18 & 23:48:20 &  56.9 &  172.4 &         26.0 &         49.0 &    H03N &     14188.7 &        319.3 &             02:29:30 &                  13.97 \\
         2010-12-25 & 23:24:00 &  38.0 &  158.0 &         26.0 &         33.0 &    H11N &      2239.6 &        340.0 &             23:50:37 &                   2.20 \\
         2004-10-07 & 13:14:43 & -27.3 &   71.5 &         35.0 &         18.0 &    H08S &      2225.9 &        182.3 &             13:41:39 &                   2.19 \\
         2004-10-07 & 13:14:43 & -27.3 &   71.5 &         35.0 &         18.0 &    H08N &      2225.9 &        182.3 &             13:41:39 &                   2.19 \\
         2010-07-06 & 23:54:43 & -34.1 & -174.5 &         26.0 &         14.0 &    H11N &      6266.1 &        161.2 &             01:06:40 &                   6.17 \\
         2010-07-06 & 23:54:43 & -34.1 & -174.5 &         26.0 &         14.0 &    H10S &     14885.6 &        203.0 &             02:43:44 &                  14.65 \\
         2010-07-08 & 23:54:45 & -34.1 & -174.5 &         26.0 &         14.0 &    H10N &     14885.6 &        203.0 &             02:43:44 &                  14.65 \\
         2016-02-06 & 13:55:09 & -30.4 &  -25.5 &         31.0 &         13.0 &    H10N &      2745.6 &        203.4 &             14:27:44 &                   2.70 \\
         2013-04-30 & 08:40:38 &  35.5 &  -30.7 &         21.2 &         10.0 &    H10S &      5129.9 &        341.5 &             09:39:32 &                   5.05 \\
         2013-04-30 & 08:40:38 &  35.5 &  -30.7 &         21.2 &         10.0 &    H10N &      5129.9 &        341.5 &             09:39:32 &                   5.05 \\
2014-08-23 & 06:29:41 & -61.7 &  132.6 &         22.2 &          7.6 &    H08S &      7759.5 &        154.0 &             07:58:15 &                   7.64 \\
         2022-02-07 & 20:06:26 & -28.7 &   11.4 &         26.5 &          7.0 &     H10N &      3548.5 &        133.8 &             20:47:49 &                   3.49 \\
         2023-04-15 & 08:22:01 & -20.1 &   36.0 &         41.4 &          6.3 &    H04N &      3262.7 &        328.3 &             09:00:59 &                   3.21 \\
         2023-04-15 & 08:22:01 & -20.1 &   36.0 &         41.4 &          6.3 &    H04S &      3262.7 &        328.3 &             09:00:59 &                   3.21 \\
         2006-04-04 & 11:30:08 &  26.6 &  -26.6 &         25.0 &          5.0 &    H10S &      4066.3 &        341.5 &             12:17:16 &                   4.00 \\
         2006-04-04 & 11:30:08 &  26.6 &  -26.6 &         25.0 &          5.0 &    H10N &      4066.3 &        341.5 &             12:17:16 &                   4.00 \\
\bottomrule
\end{tabular}}
\caption{A summary of the most energetic fireballs occurring over open ocean and the associated hydroacoustic stations for which a direct signal path is possible. Here we have examined all bolides with energies above 5 kT from the CNEOS list. Note that each row corresponds to a possible signal path to a station where data was available; some fireballs have more than one potential station where a signal may be detected. The date and time of the fireball as well as its latitude, longitude and height of peak brightness are shown together with the total energy of the event in kilotons of TNT (1 kT = 2.184$\times 10^{12}$ J). The range (R) from the fireball to each station is given as well as the expected azimuth of arrival (Az) at the station, the nominal arrival time (in UT) assuming a celerity of 1.48 km/s (and allowing for travel time from the height of peak brightness for either shock waves or large meteorites). The final column, Dur, is the size of the time window (in minutes) during which a signal might be detected, presuming celerities range from 1.42 - 1.55 km/s.}
\label{tab:cneos_high_energy}
\end{table}
\end{adjustwidth}
\end{landscape}

\begin{landscape}
\begin{table}[h!]
\centering
\begin{tabular}{llrrrrlrrlr}
\toprule
               Date &     Time (UT) &   Lat &   Long &  Ht (km) &  E (kT) & Station &  R(km) &  Az & Arr. Time &  Dur \\
\midrule
2005-10-26 &  21:30:47 & -36.3 & -80.5 &         16.7 &        0.400 &    H03N &         358 &        206.8 &             21:35:29 &                  0.33 \\
2015-02-17 &  13:19:50 &  -8.0 & -11.2 &         39.0 &        0.110 &    H10N &         352 &         90.2 &             13:25:54 &                  0.35 \\
2008-12-12 &  11:36:36 &  -7.0 &  -9.7 &         50.0 &        0.600 &    H10N &         530 &         78.2 &             11:45:15 &                  0.52 \\
2008-12-12 &  11:36:36 &  -7.0 &  -9.7 &         50.0 &        0.600 &    H10S &         530 &         78.2 &             11:45:15 &                  0.52 \\
2022-09-04 &  03:54:55 &  20.0 & 165.9 &         21.9 &        0.086 &    H11N &         106 &        316.8 &             03:57:17 &                  0.10 \\
2025-03-13 &  09:46:51 &  -7.1 &  76.0 &         38.0 &        1.800 &    H08S &         398 &         87.0 &             09:53:22 &                  0.39 \\
2022-03-28 &  10:20:24 &  -7.7 &  74.3 &         29.0 &        0.980 &    H08S &         214 &        102.1 &             10:24:22 &                  0.21 \\
2012-09-28 &  05:44:12 &  -6.9 &  73.7 &         25.0 &        0.130 &    H08S &         150 &         73.0 &             05:47:14 &                  0.15 \\
2012-09-28 &  05:44:12 &  -6.9 &  73.7 &         26.0 &        0.130 &    H08N &         150 &         73.0 &             05:47:14 &                  0.15 \\
\bottomrule
\end{tabular}

\caption{A summary of fireballs occurring within 500 km range of hydroacoustic stations where data and a direct propagation path exist.  Here we have examined all bolides down to the CNEOS energy limit. The columns are the same as Table \ref{tab:cneos_high_energy}.}
\label{tab:cneos_near_stations}
\end{table}
\end{landscape}

\begin{landscape}
\begin{adjustwidth}{-1.5cm}{-1.5cm}
\begin{table}[h!]
\centering
{\renewcommand{\arraystretch}{0.6}
\begin{tabular}{llrrrrlrrlr}
\toprule
               Date &     Time (UT)&   Lat &   Long &  Ht (km) &  E (kT) & Station &  R(km) &  Az & Arr. Time &  Dur \\
\midrule
2004-09-03 &  12:07:22 & -67.7 &   18.2 &         31.5 &         13.0 &    H01W &        6795 &        205.5 &             13:25:35 &                   6.7 \\
2003-12-09 &  22:36:23 & -67.1 &  -90.8 &         25.0 &          1.6 &    H03S &        3806 &        188.3 &             23:20:35 &                   3.8 \\
2003-12-09 &  22:36:23 & -67.1 &  -90.8 &         26.0 &          1.6 &    H03N &        3806 &        188.3 &             23:20:35 &                   3.8 \\
2010-09-03 &  12:04:58 & -61.0 &  146.7 &         33.3 &              &    H01W &        3720 &        152.6 &             12:48:38 &                   3.7 \\
2010-09-03 &  12:04:58 & -61.0 &  146.7 &         34.3 &              &    H08S &        8455 &        151.3 &             13:41:58 &                   8.3 \\
2010-09-03 &  12:04:58 & -61.0 &  146.7 &         35.3 &              &    H08N &        8455 &        151.3 &             13:41:58 &                   8.3 \\
2014-08-23 &  06:29:41 & -61.7 &  132.6 &         22.2 &          7.6 &    H01W &        3286 &        163.0 &             07:07:53 &                   3.2 \\
2014-08-23 &  06:29:41 & -61.7 &  132.6 &         23.2 &          7.6 &    H08S &        7759 &        154.0 &             07:58:15 &                   7.6 \\
2017-06-20 &  13:41:32 & -54.2 &  133.0 &         33.3 &          1.6 &    H01W &        2611 &        153.3 &             14:12:43 &                   2.6 \\
2017-06-20 &  13:41:32 & -54.2 &  133.0 &         34.3 &          1.6 &    H04N &        5501 &        130.5 &             14:45:16 &                   5.4 \\
2017-06-20 &  13:41:32 & -54.2 &  133.0 &         35.3 &          1.6 &    H04S &        5501 &        130.5 &             14:45:16 &                   5.4 \\
2017-06-20 &  13:41:32 & -54.2 &  133.0 &         36.3 &          1.6 &    H08S &        7470 &        146.4 &             15:07:26 &                   7.3 \\
2022-01-11 &  03:33:13 & -58.4 & -160.2 &         40.8 &          2.9 &    H01W &        6518 &        142.0 &             04:48:48 &                   6.4 \\
2022-01-11 &  03:33:13 & -58.4 & -160.2 &         41.8 &          2.9 &    H03S &        6400 &        218.0 &             04:47:28 &                   6.3 \\
2020-03-04 &  20:25:59 & -53.5 &   90.8 &         24.3 &          1.0 &    H01W &        2866 &        214.4 &             20:59:25 &                   2.8 \\
2020-03-04 &  20:25:59 & -53.5 &   90.8 &         25.3 &          1.0 &    H04N &        2853 &        120.4 &             20:59:25 &                   2.8 \\
2020-03-04 &  20:25:59 & -53.5 &   90.8 &         26.3 &          1.0 &    H04S &        2853 &        120.4 &             20:59:25 &                   2.8 \\
2020-03-04 &  20:25:59 & -53.5 &   90.8 &         27.3 &          1.0 &    H08S &        5400 &        165.5 &             21:28:04 &                   5.3 \\
2013-07-30 &  02:36:58 & -50.2 &   90.2 &         25.6 &          1.0 &    H01W &        2690 &        221.0 &             03:08:38 &                   2.6 \\
2013-07-30 &  02:36:58 & -50.2 &   90.2 &         26.6 &          1.0 &    H08S &        5050 &        164.0 &             03:35:12 &                   5.0 \\
2003-11-10 &  13:54:06 & -64.5 &  136.2 &         23.0 &          1.3 &    H01W &        3650 &        163.0 &             14:36:24 &                   3.6 \\
2003-11-10 &  13:54:06 & -64.5 &  136.2 &         24.0 &          1.3 &    H04S &        5220 &        144.0 &             14:54:05 &                   5.1 \\
2003-11-10 &  13:54:06 & -64.5 &  136.2 &         25.0 &          1.3 &    H08S &        8050 &        156.0 &             15:25:56 &                   7.9 \\
2003-09-02 &  20:00:46 &  51.3 & -161.0 &         25.0 &          1.0 &    H03N &       12362 &        318.4 &             22:21:19 &                  12.2 \\
2021-07-05 &  03:46:24 &  44.3 & -164.2 &         43.4 &          1.8 &    H03N &       12209 &        310.7 &             06:06:13 &                  12.0 \\
2021-07-05 &  03:46:24 &  44.3 & -164.2 &         44.4 &          1.8 &    H03S &       12209 &        310.7 &             06:06:13 &                  12.0 \\
2021-07-05 &  03:46:24 &  44.3 & -164.2 &         45.4 &          1.8 &    H11N &        3877 &         37.6 &             04:32:23 &                   3.8 \\
2006-01-28 &  03:33:48 & -51.7 &   56.4 &         37.0 &          1.8 &    H01W &        5000 &        228.6 &             04:32:06 &                   4.9 \\
\bottomrule
\end{tabular}}

\caption{A summary of fireballs occurring over open ocean where the SOFAR channel depth is predicted to be less than 500m from the surface. Only hydroacoustic stations where data and a direct propagation path exist are shown.  Here we have examined all bolides down to an energy limit of 1 kT (or where no energy is given). The columns are the same as Table \ref{tab:cneos_high_energy}.}
\label{tab:cneos_shallow_SOFAR}
\end{table}
\end{adjustwidth}
\end{landscape}

\section{Discussion}\label{discussion}

We have described evidence of a possible direct (H-phase) hydroacoustic signal detected on H03N from a fireball which occurred Sep 2, 2003 off the coast of Alaska. However, as only one station detected this signal, the possibility of a false positive (random) association needs to be estimated.

To perform this estimation, we assume that all non-repeating hydroacoustic signals which we identified during our survey occur at random azimuths. Within the total of all 53h of our survey hourly search ``blocks", we find an average family detection rate of one per 17 minutes. Given our acceptance azimuth of $\pm$ 5$^{\circ}$ and assuming Poisson statistics we find that the chances that any one of the 15 families which occur within our acceptance time window will match a fireball azimuth by chance is 27\%. This high level of potential false positive indicates that we cannot connect the signal with the fireball in a statistical sense as anything more than a weak association. 

Given that our assumption of random arrival directions is likely too simple, we also checked the false positive rate by shifting arrival windows by 5 min across each one hour time block per station. We found the resulting associations varied between 0 and 3 for the ten time blocks used as a check, with an average near one, further suggesting that our one lone association is likely a false positive.

If we assume the H03N signal is related to the Sep 2, 2003 fireball, we can also estimate the rough order of magnitude of the hydroacoustic energy coupling efficiency, $\eta$, to compare with predictions. For this we appeal to Equation \ref{eq:energy} which relates source yield in the SOFAR channel and hydroacoustic pressure amplitude as a function of range. Using this equation, our observed hydroacoustic pressure of 0.04 Pa, the known range from the fireball (12,300 km) to H03N and the estimate for the fireball energy (1 kT) we arrive at a coupling efficiency $\eta$ = 5$\times 10^{-9}$. 

For comparison, \cite{Clarke1996} provides a theoretical estimate of the hydroacoustic coupling efficiency for near surface nuclear bursts using the CALE hydrodynamics code. They found that for 1 kT nuclear detonation at an altitude of 1 km above the ocean has an $\eta$ $\sim$ 10$^{-7}$. Note that this efficiency refers to the energy remaining at 10 km range, where \citet{Clarke1996} assumes that the energy is then effectively trapped in the SOFAR channel and hence appropriate for deep ocean, long range propagation estimates. At a typical fireball altitude of 25 km this would be expected to be even lower as shock energy is further removed as altitude increases \citep{Glasstone1977, Collins2017}, broadly consistent with our result.

More generally, we have 54 total stations where a direct-path fireball signal could be expected to be detected (but wasn't in at least 53). We may use these non-detections to set an upper limit as to the hydroacoustic coupling efficiency of fireballs over the open ocean.

To estimate the minimum signal detectable at each station, we use our DTK-PMCC list of families and their associated peak amplitude. For each station, we use all the families in the 1h search window to estimate the minimum peak pressure required to produce a detection at that station in the conditions at the time of the expected fireball signal arrival. Using the known range to each fireball (and its energy), this estimated minimum detectable pressure is combined with equation \ref{eq:energy} to produce an equivalent minimum energy which would be detectable in the SOFAR channel. 

We caution that this approach is not meant to reflect a rigorous, formal metric in a detection-theory sense, as minimum detectable amplitudes vary with noise conditions and with choice of window lengths used and may also be expected to be a function of frequency band, coherence and choice of slowness mismatch chosen for family/pixel detection in DTK-GPMCC. As a result, the values serve only to provide a more stringent upper bound to the coupling coefficient estimation and is strictly only appropriate to our chosen analysis pipeline. 

A better procedure would be to inject synthetic signals to test the detection sensitivity of our chosen parameters for family association using DTK-GPMCC. Such an approach would require producing a source function, which could in future be computed on a per event basis using a hydrocode approach similar to the work of \citet{Clarke1996} and fully tracking the resulting waveform signal into and through the SOFAR channel. This would allow confidence bounds to be placed on our coupling estimates, rather than our simple approach which produces approximate upper bounds.

This SOFAR explosive yield limit when combined with the fireball energy produces an upper limit to the coupling efficiency. Across all 53 stations using this noise background estimation we find an average upper limit to the airburst-to-SOFAR coupling efficiency of 10$^{-10}$. It is worth nothing that for some of the kT-class fireballs we examined, this limit is as low as 10$^{-14}$.

We again emphasize that the coupling coefficient derived are order of magnitude at best. Our analysis assumes only straight path propagation and ignores reflections and diffraction from undersea features as well as out of plane refraction effects. The results are entirely conditional on our simplified geometry and the frequency band (0.1 – 40 Hz)  used in our GPMCC detection algorithm. A better approach would be to perform detailed modeling using hydrocode to produce a realistic shock \citep{Shuvalov2013, Avramenko2014, Collins2017, Aftosmis2016} per event and then accurately track the energy partitioning across the entire air-ocean ground footprint and with depth into a stratified ocean and through the SOFAR channel. This is well beyond the scope of this work.

For comparison, for the solid Earth case, \cite{Svetsov2019} finds a theoretical seismo-acoustic coupling efficiency for shocks from larger airbursts to be within a factor of two of 2.5$\times10^{-5}$. Observational studies of smaller meteors \citep{Edwards2008} suggest an even larger range from 10$^{-2} \leq \eta \leq 10^{-7}$ albeit at much smaller energies. These variations were attributed by \citet{Edwards2008} to differing local soil conditions near individual seismic stations. 

These efficiencies are much higher than we find for SOFAR-coupling, but it should be noted that the comparison would be more appropriate if we could measured the energy just below the water surface. Instead, we can only say that depositing energy from acoustic waves or impacts at the ocean surface even to shallow SOFAR channels appears to be very inefficient.

As a final check to confirm if other processes might be masking the direct hydroacoustic signal from fireballs, we note that the handful of previous hydroacoustic studies which examine explosive sources emphasize the importance of underwater reflections \citep[e.g.]{Vergoz2021}. This is due to the sometimes complex underwater bathymetry which can attenuate propagation in the SOFAR channel \citep{Lin2025}, while enhancing it in other directions through reflections. 

We explored the possibility that for the most energetic events (E $\geq$ 10 kT) signals from reflections might be produced. The details of this exploration can be found in the appendices, where one case study is presented in detail. The summary of our reflection exploration is that several potential reflected signals were found in extended time windows around the expected arrival times for four large fireballs. These were reflections where the arrival timing and backazimuth were explainable as being reflections off underwater features protruding into the SOFAR channel. However, while we could find plausible the role of reflections the lack of direct arrivals in most cases is puzzling. Moreover the waveforms are more similar to Earthquake T-phases than explosive sources. For example, a potential reflected signal for the Feb 7, 2002 shows what appear to be clearly separated S and P wave arrivals, almost certainly related to the signal being produced by a local Earthquake. 

The high efficiency of sonicfication of Earthquakes into the SOFAR channel \citep{Okal_2008} implies that very small magnitude Earthquakes can produce significant signals. None of the reflected events are associated with documented Earthquakes in catalogues, but this is not surprising given the high efficiency of T-phase production from small Earthquakes.

We conclude that none of these potential signals is convincingly associated with any of the energetic fireballs. More likely these are false positives, particularly given the broad acceptance criteria when using the numerous underwater reflectors. More details are summarized in the appendices. 

Of the the most energetic fireballs recorded over the ocean, the Bering Sea fireball of Dec 18, 2018 stands out with a yield near 50 kT. It had an expected arrival at the only station (H11) where a direct propagation path was not blocked by land at the same time as a nearby earthquake which produced an extended signal, potentially masking any arrivals. 

The large South Sulawesi fireball on Oct 8, 2009 \citep{Silber2011} occurred over the Gulf of Boni, Indonesia releasing 33 kT of energy. Unfortunately all propagation paths to the open ocean are blocked by islands; hence it is unsurprising that this bolide was not detected by any IMS hydroacoustic stations.   

The low coupling efficiencies we find for these long range hydroacoustic detections suggest open ocean airbursts are much easier to detect infrasonically or seismically. Our study suggests that even low altitude detonations of order a few to 10 kT would be challenging to detect through shock-coupling.  Our lack of success in detecting hydroacoustic signals from bolides with high energies, those occurring at shallow SOFAR locations or near hydroacoustic stations suggests detecting explosions over the open ocean is best done through other technologies.

\section{Conclusion}

The main finding in this work is that direct hydroacoustic signals from fireballs are very rare. We set an upper limit for the efficiency of high altitude fireball shock production to couple into the SOFAR channel of $\eta \leq 10^{-10}$. We emphasize that this efficiency is conditional on our chosen set of DTK-GPMCC signal processing settings, assumes a direct path H-phase signal and a signal celerity range of 1.42-1.55 km/s. Within these assumptions we find no unambiguous detections among 53 station-fireball pairs. Using the yield-range-pressure relation of Eq. \ref{eq:energy} we find SOFAR-equivalent yields by assuming the minimum detectable amplitude signal family associations in our data reflects the noise background across our survey to arrive at this upper efficiency limit estimate. Confirming a positive detection with more than one station is complicated by the lack of complete hydroacoustic stations, as loss of cabling or a single element can take years to fix given the cost and complexity of station replacement \citep{Bittner2024}. This means some stations have long time periods without data or with only two elements precluding backazimuth estimation and making multi-station detection very rare. Complex propagation paths and land blockage also reduce coverage.

While meteorite fragments impacting the ocean surface may (in unusually energetic cases of large fragments) be able to produce a signal, the shockwave from a bolide is more likely to be detected by hydroacoustic stations in the SOFAR channel. This conclusion is based on energy considerations. In this scenario the fireball airshocks are transmitted at the ocean surface and then propagate downwards to the SOFAR channel where a tiny fraction of the original energy is trapped, permitting long distance propagation. 

One possible hydroacoustic fireball event we identified (Sep 2, 2003), if real, would suggest a coupling efficiency for hydroacoustic production in the SOFAR channel can be of order 10$^{-9}$, using the amplitude-range and in-ocean source energy relations of \cite{Soloway2014}. However, the high statistical probability of this being a false positive suggests this is more likely a chance association. This suggests that low altitude explosions with energies of order a few kT would be very difficult to detect by hydrophones at oceanic distances. The largest hurdle to H-phase detection appears to be the difficulty in transmitting energy from the shock at the ocean surface to the SOFAR channel. 

The lack of hydroacoustic fireball detections emphasizes the role for infrasound and seismic technologies in characterizing bolides over the ocean. Both of these techniques have demonstrated capability in detecting  fireball shock waves based on studies over many decades \citep{Silber2019, Edwards2008}.

\section*{Acknowledgments}
Funding for this work was provided in part through NASA co-operative agreement 80NSSC24M0060, from the Natural Sciences and Engineering Research Council of Canada, the Canada Research Chairs program and Natural Resources Canada.


%
%



\bibliographystyle{cas-model2-names} 
\bibliography{bibliography.bib}

%
%
%
%
%

\appendix
\clearpage
\section{Real world explosive comparisons of yield-range-pressure relation.}
\label{shot_cals}

The data used for Equation \ref{eq:energy} is drawn mainly from archival measurements by \citet{Arons_1954}. The expression was derived for explosions occurring in the SOFAR channel and then measured at large ranges. To attempt a modern validation of this expression using our signal processing chain we searched the literature for explosions of known yield, timing and location during the epoch when IMS hydroacoustic stations were recording data. 

The results are summarized in Table \ref{tab:hydro-calibration}. Unfortunately, we were not able to find any modern explosive shots which took place at depth in the SOFAR channel with all known quantities also published. As a result, we may only check the relation in comparison to the existing literature which exclusively is limited to shallow explosions, typically only a few tens of meters below the ocean surface.

Using Eq. \ref{eq:energy} and comparing the measured pressure to that expected, we consistently find that the relation overpredicts pressure for an equivalent yield in the SOFAR channel. We interpret this as an indication that for explosions outside the SOFAR channel only a small portion of the energy leaks into the SOFAR channel to be detected at large range. As noted by \citet{Clarke1996}, explosions in shallow water and the particulars of reflection from the ocean bed can dramatically change both the spectrum and magnitude of the pressure measured at long distances in the SOFAR channel. \citet{Prior2010, Hanson_Reasoner_2007} also note that the depth of an explosion has a large influence on the resulting yield estimation using distant pressure measurements, a well known result in hydroacoustics \citep[e.g.][]{Gao2022}. As none of these explosions are in the SOFAR channel, we cannot explicitly validate Eq. \ref{eq:energy}.  

\begin{landscape}
\scriptsize
\renewcommand{\arraystretch}{1.05}
\setlength{\tabcolsep}{3pt}
\setlength{\LTleft}{-1.5cm}
\setlength{\LTright}{-1.5cm}

\begin{longtable}{
>{\raggedright\arraybackslash}p{1.45cm}
>{\centering\arraybackslash}p{1.45cm}
>{\centering\arraybackslash}p{0.85cm}
>{\centering\arraybackslash}p{1.05cm}
>{\centering\arraybackslash}p{0.85cm}
>{\centering\arraybackslash}p{1.20cm}
>{\centering\arraybackslash}p{1.00cm}
>{\centering\arraybackslash}p{1.45cm}
>{\centering\arraybackslash}p{1.35cm}
>{\centering\arraybackslash}p{1.70cm}
>{\centering\arraybackslash}p{1.35cm}
>{\centering\arraybackslash}p{1.90cm}
>{\centering\arraybackslash}p{4.60cm}
}
\caption{Literature cases where underwater explosions of known yield, location and time produce signals detected at IMS Hydroacoustic stations. The arrival time and amplitude were found using the same DTK-GPMCC parameters adopted for our fireball survey. The inferred yield is the estimated SOFAR equivalent explosive energy required to produce the observed pressure at the known range according to Eq. \ref{eq:energy}.}\label{tab:hydro-calibration}\\
\toprule
\makecell[l]{Date} &
\makecell[c]{Origin Time\\(UTC)} &
\makecell[c]{Lat} &
\makecell[c]{Long} &
\makecell[c]{Station} &
\makecell[c]{Arrival\\Time} &
\makecell[c]{BackAz} &
\makecell[c]{Yield\\(kg TNT)} &
\makecell[c]{Distance\\(m)} &
\makecell[c]{Amplitude\\(Pa)} &
\makecell[c]{Inferred Yield\\(kg TNT)} &
\makecell[c]{Reference} &
\makecell[c]{Comments} \\
\midrule
\endfirsthead

\toprule
\makecell[l]{Date} &
\makecell[c]{Origin Time\\(UTC)} &
\makecell[c]{Lat} &
\makecell[c]{Long} &
\makecell[c]{Station} &
\makecell[c]{Arrival\\Time} &
\makecell[c]{BackAz} &
\makecell[c]{Yield\\(kg TNT)} &
\makecell[c]{Distance\\(m)} &
\makecell[c]{Amplitude\\(Pa)} &
\makecell[c]{Inferred Yield\\(kg TNT)} &
\makecell[c]{Reference} &
\makecell[c]{Comments} \\
\midrule
\endhead

\midrule
\multicolumn{13}{r}{Continued on next page} \\
\endfoot

\bottomrule
\endlastfoot

2008-09-06 &  &  &  & H11N &  & 312 & 39 & 3000000 & 1.4 & 0.25 & Prior et al., 2011 & average of 36 returns from 6:15 - 23:44 UT with Az 310-313 \\
2008-09-07 &  &  &  & H11N &  & 314 & 39 & 3000000 & 1.85 & 0.5 &  & average of 19 returns from 6:00 - 09:44 UT with Az 313-315 \\
 &  &  &  & H03N &  & 287 & 39 & 16300000 & 1.2 & 22 &  & average of 16 returns from 6:00-12:07 UT with Az 286-288 \\
2008-09-08 &  &  &  & H11N &  & 315 & 39 & 3000000 & 1.8 & 0.45 &  & average of 13 returns from 6:00-09:40 UT with Az 313-317 \\
 &  &  &  & H03N &  & 288 & 39 & 16300000 & 0.7 & 5 &  & average of 12 returns from 6:00-09:45 UT with Az 288-289 \\
2008-09-09 &  &  &  & H11N &  & 315 & 39 & 3000000 & 0.9 & 0.06 &  & average of 5 returns from 6:00-08:20 UT with Az 313-318 \\
 &  &  &  & H03N &  & 288 & 39 & 16300000 & 0.18 & 0.15 &  & average of 6 returns from 6:00-09:12 UT with Az 288-289 \\
2016-06-10 & 17:10:48.97 & 29.94 & -79.58 & H10N & 18:41:21 & 304.7 & 6759 & 8128000 & 13.3 & 1650 & Heyburn et al., 2018 & Observed Backazimuth of 304.7 \\
2016-06-23 & 17:20:00 & 29.95 & -79.85 & H10N & 18:50:52 & 304.5 & 6759 & 8128000 & 13.5 & 1750 &  & Observed Backazimuth of 305.1 \\
2016-07-16 & 20:00:12.28 & 29.68 & -79.57 & H10N & 21:30:34 & 304.4 & 6759 & 8177000 & 9.3 & 650 &  & Observed Backazimuth of 304.6 \\
2016-09-04 & 18:29:31.68 & 30.362 & -79.537 & H10N & 20:00:00 & 304.5 & 6759 & 8177000 & 15.1 & 2325 &  & Observed Backazimuth of 304.2 \\
2016-09-21 & 16:30:54.83 & 30.173 & -79.568 & H10N & 18:01:30 & 304.5 & 6759 & 8177000 & 10.05 & 800 &  & Observed Backazimuth of 303.8 \\
2017-12-01 & 20:04:55.19 & -45.65 & -59.4 & H10N & 21:11:58 & 217.7 & 102 & 5995000 & 0.91 & 0.5 & Nielsen et al., 2021 & Observed Backazimuth of 217.1 \\
 &  &  &  & H04S & 21:32:49 & 223.9 &  & 7778900 & 0.22 & 0.04 &  & Observed Backazimuth of 225.3 \\
2021-06-18 & 19:48:40 & 29.7167 & -79.4667 & H10N & 21:19:14 & 304.5 & 18000 & 8109000 & 39.6 & 30000 & Bittner et al., 2024 & Observed Backazimuth of 304.4 \\
2021-07-16 & 17:48:34 & 30.0998 & -79.482 & H10N & 19:19:01 &  & 18000 & 8109000 & 19.2 & 4350 & David Dall'Osto & Observed Backazimuth of 304.4 \\
 &  &  &  &  &  &  &  &  &  &  & Personal Communication &  \\
 &  &  &  &  &  &  &  &  &  &  &  &  \\
2021-08-08 & 19:55:18 & 30.1197 & -79.7057 & H10N & 21:26:18 & 304.9 & 18000 & 8146000 & 15.2 & 2325 & David Dall'Osto & Observed Backazimuth of 304.7 \\
 &  &  &  &  &  &  &  &  &  &  & Personal Communication&  \\
 &  &  &  &  &  &  &  &  &  &  & &  \\

\end{longtable}
\end{landscape}

\clearpage
\section{Exploring the Importance of Hydroacoustic Reflectors} \label{reflection_section}

As discussed in the main text, our initial examination of possible signals showed that direct arrivals for all but one fireball were not present. While this most likely implies that fireball produced hydroacoustic signals are rare/difficult to produce, we also explored the possibility that signals for large events might be produced but commonly blocked due to underwater topography. In some cases, these signals might be visible as reflections, as hydroacoustic reflections are in fact common at moderate to large ranges \citep[e.g][]{Pulli1999, Tolstoy2006}, though direct arrivals are usually also present.

For two of the highest energy fireballs a significant signal was found at a station within the 1h search time window, but the temporal and backazimuth of the signal were inconsistent with a direct arrival.  Upon further examination we found a few of these signals to be temporally and spatially consistent with a reflection off a plausible bathymetric feature that extended into the SOFAR channel.  We have identified these as possible indirect arrivals. We include these where they allow temporal and backazimuthal correspondence consistent with a fireball hydroacoustic source when a reflection region is included in the analysis. 

To search for such reflections, we followed the methodology of \cite{Tolstoy2006,Vergoz2021}, and isolate signals where the timing and the backazimuths obtained at the station were high enough signal-to-noise (SNR) to find a suitable sub-surface reflector. 

As described in the main text, using this approach, for each prospective hydroacoustic arrival in our acceptance window, the backazimuths obtained through DTK-PMCC are employed to extend the propagation path to the geometric coordinates of the meteor source. If the projected backazimuth intersects the fireball location from CNEOS estimates to within an azimuthal tolerance of $\sim5^{\circ}$ through a water path that extends in depth to the height of the SOFAR channel without seamount interruptions, then we designate this path as a likely direct arrival. As summarized in the main text, only the Sep 2, 2003 fireball off the Alaskan coast met this criteria - no other direct arrivals from any of the 29 other fireballs were found.

Alternatively, if there was an underwater feature above the depth of the SOFAR channel, or the backazimuth did not point directly to the source, then we considered the possibility that the acoustic path reflected off an underwater feature. Such a reflection would occur after the acceptance window but still inside the 1h search window, given the typical distances involved. 

In order to find potential reflected signals we manually examined signals occurring after the acceptance window but before the end of the 1h search window. We computed the expected travel time from the source to potential reflecting points and then to the hydrophone. We manually located likely bathymetric reflectors based on the signal backazimuth. 

This was entirely a manual search process meant to see if any plausible reflecting signals could be found for the most energetic fireball population. We understood a priori that any plausible signals would necessarily be suspect as potential false positives and would have to be individually examined to see if T-phase Earthquake contamination could be ruled out.

For a suspected reflection, we would define three locations:

\begin{enumerate}
    \item Source - The point of the meteor source, given by USG, corresponding to the approximate brightest point along the trajectory
    \item Receiver - The location of the hydroacoustic array, specifically the location given by CTBTO
    \item Reflector - A bathymetric feature above the lower depth of the SOFAR channel that forms a possible reflector given the observed backazimuth at the hydrophone. For simplicity, following \cite{Vergoz2021} we only assume one reflector per path.
\end{enumerate}

Possible reflectors were found using a grid search, where triangles were drawn between the receiver, source, and the potential reflector represented as a grid point, using the backazimuth as a constraint within our $5^{\circ}$ azimuth tolerance. This results in a region of potential reflection points consistent with the travel time from the source. 

These points were further temporally constrained by ray-tracing, as discussed in Section \ref{timing_section}. In this manner, we could pick a tolerance based on the distance and the total travel time which was set to be on the order of a few minutes, given uncertainties in the time of excitation for a SOFAR signal source from the fireball acoustic wave. We then examined a map of the local bathymetry in the potential reflection region, filtering out locations where the ocean bottom depth was below the SOFAR channel. We then compared our grid search results to potential reflection zones based on this bathymetry. If there existed a region where the ocean bottom penetrated to within the SOFAR channel within our grid search region, then that point was tagged as a potential reflection point in our study.

We emphasize that this process was designed to provide limits only on the frequency of possible reflected signals; if none could be found this would provide yet another limit on hydroacoustic production from fireballs. Conversely, if some plausible reflections were found we still need to examine the signal characteristics, direct and reflected transmission propagation paths to further understand if these are real signals a randomly associated background signal.

\subsection{Propagation and Timing} \label{timing_section}

The timing and propagation attenuation was calculated using PyRAM\footnote{https://github.com/marcuskd/pyram}. The PyRAM solver is a Python wrapper of the RAM solver (Range-dependent Acoustic Model) \citep{Collins1999}, which calculates hydroacoustic transmission loss and travel time between source and receiver, given 2 dimensional sound speed and bathymetry profiles. The inputs for these programs are bathymetry and range and depth dependent sound speed. Ray-tracing was conducted in 2D searches, where the reflection searches described in \ref{reflection_section} were effectively "2D $\times$ N" instead of 3D, using a series of 2D searches and changing the launch angle to simulate a 3D search, ignoring interaction effects. The ray-path timing was calculated assuming straight line propagation between any two points.  For these reflected arrivals, we would calculate the time between source and reflector, and then add the time between reflector and receiver. For each pair, we describe the start and end points used as initial and final.

The bathymetry model data was  from the National Oceanagraphic and Atmospheric Adminstrations (NOAA), Topography, SRTM30+ Version 6.0, using 30 arc second resolution\footnote{https://coastwatch.pfeg.noaa.gov/erddap/griddap/usgsCeSrtm30v6.html} for the grid. We found the geographic coordinates between the initial and final points using the Pyproj\footnote{https://pyproj4.github.io/pyproj/stable/} library, and sliced the grid into a line bathymetry profile.

For sound speed, the salinity and potential temperature data was extracted from the National Centers for Environmental Prediction (NCEP) Global Ocean Data Assimilation System (GODAS)\footnote{https://psl.noaa.gov/data/gridded/data.godas.html}. We sliced the gridded data following the same procedure as the bathymetric data. The potential temperature, $\theta$, was converted into temperature using the definition of potential temperature \citep{Talley2011}:

\begin{equation}
    \theta = T \left(\frac{p_0}{p} \right)^{\frac
    {R}{c_p}}
\end{equation}
where, $T$ is the temperature of the water in degrees Celsius, $p_0$ is the reference pressure at the surface, $p$ is the pressure of the fluid at the depth of the measurement, $R$ is the ideal gas constant and $c_p$ is the specific heat capacity at constant pressure for water ($4184$ J/kg/K). \par

We convert temperature and salinity to sound speed, $c$, using the empirical equation taken from \cite{Mackenzie1981}:
\begin{multline}
    c = 1448.96 + 4.591T - 5.304 \times 10^{-2}T^2 + 2.374\times 10^{-4}T^3 \\ + 1.340(S - 35) + 1.630 \times 10^{-2}z + 1.675 \times 10^{-7}z^2 - 1.025\times 10^{-2}T(S - 35)\\ - 7.139 \times 10^{-13}Tz^3
\end{multline}

where $s$ is salinity in g/kg, $z$ in the depth in meters and $T$ is temperature in Kelvin. \par

\section{Events showing possible reflections}

There were two fireballs among our high energy population where we found possible late arriving reflected hydroacoustic signals based on timing and backazimuth reflection from a plausible underwater feature. We describe the potential reflected hydroacoustic signals in detail for the most promising of the two (Oct 7, 2004), noting that the other one (Dec 25, 2010) the conclusions are basically the same (ie the reflections are likely false associations produced from Earth quake T-phases).

The complex scattering, shadowing and reflection geometries we find for hydroacoustic signal propagation for these fireball events, mirror similar scenarios proposed in associating hydroacoustic signals with underwater submarine explosions  \citep{Vergoz2021, Nielsen2021} and documented for controlled explosions \citep{Pulli1999}. Similar effects found by  \citet{Heaney2013} for an undersea volcanic eruption localization study also emphasized the shortcomings in using 2D propagation tracing for association.

\subsection{October 7, 2004}


\begin{table}[]
\resizebox{\textwidth}{!}{\begin{tabular}{@{}llllllllllllll@{}}
\toprule
\begin{tabular}[c]{@{}l@{}}Arrival \\ No\end{tabular} & Station & \begin{tabular}[c]{@{}l@{}}Time of \\ Arrival\\ {[}UTC{]}\end{tabular} & \begin{tabular}[c]{@{}l@{}}Difference from\\ Expected Time\\ {[}s{]}\end{tabular}  & Observed? & \begin{tabular}[c]{@{}l@{}}Direct/\\ Indirect?\end{tabular} & \begin{tabular}[c]{@{}l@{}}Distance \\ {[}km{]}\end{tabular} & \begin{tabular}[c]{@{}l@{}}Travel \\ Time {[}min{]}\end{tabular} & \begin{tabular}[c]{@{}l@{}}Observed Back \\ Azimuth\\ {[}deg{]}\end{tabular} & \begin{tabular}[c]{@{}l@{}}Peak Amplitude\\ {[}Pa{]}\end{tabular} & Celerity [m/s] & \begin{tabular}[c]{@{}l@{}} Dominant \\ Frequency [Hz]\end{tabular}  & Signal to Noise [dB] & Correlation\\ \midrule
1 & H08S & 13:38:25 & - & No & Direct & 2133 & 23.7 &  & & & &&\\
2 & H08S & 14:04:57 & 0* & Yes & Indirect & 4248 & 47.2 & 68.96 & 3 & 1409 & 5.1 & 22 & 0.80\\
\\\bottomrule
\end{tabular}}
\caption{A summary of the expected direct arrivals and indirect arrival for the October 7, 2004 event at H08S. The dominant signal frequency and the signal to noise ratio is given per arrival as well as the PMCC cross-correlation value. *Location of reflector was constrained by the arrival time, so expected time will always be the same as reflected time.}
\label{2004_data_table}
\end{table}

The best candidate for a fireball-associated reflected hydroacoustic signal based purely on timing and backazmiuth considerations relative to underwater features, is a CNEOS fireball reported as occurring on Oct 7, 2004. The CNEOS data provide an estimated total energy of 18 kT based on the light production for this event which occurred about halfway between Madagascar and the western coast of Australia. The event was also detected infrasonically  \citep{Arrowsmith2007} at half a dozen IMS infrasound stations at ranges up to 18000~km. The resulting infrasound waveforms showed a multi station average period of 13 sec, which using the period-yield relation described by \cite{Ens2012} produces an expected yield near 23 kT, consistent with the optical lightcurve.  The fireball source location is approximately 2200 km away from the H08 hydroacoustic sensor pair of arrays. The expected direct arrival time at that array is 10/7/2004 13:38:25 UTC with an acceptance window of 2.2 minutes. As summarized in the main text no direct signals with the proper timing and direction were recorded at H08.

\begin{figure*}
  \includegraphics[width=\linewidth]{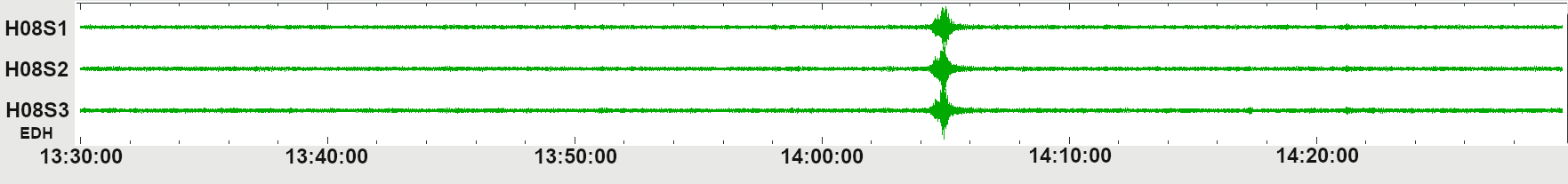}
  \caption{The hydroacoustic signal associated with the Oct 7, 2004 fireball as recorded at H08S. Shown are the unfiltered time series of underwater pressure from each of the three hydrophones beginning at 13:30 UT, Oct 7, 2004 and lasting for one hour. There is no other event of comparable magnitude in this hour window. The EDH channel code represents E - Extremely Short Period (sampling rate of about 80 - 250 Hz), D - Pressure measurements, H - Hydrophone. For more information, see https://ds.iris.edu/ds/nodes/dmc/data/formats/seed-channel-naming/}
  \label{fig:2004_H08}
\end{figure*}

\begin{figure*}[!ht]
  \includegraphics[width=\linewidth]{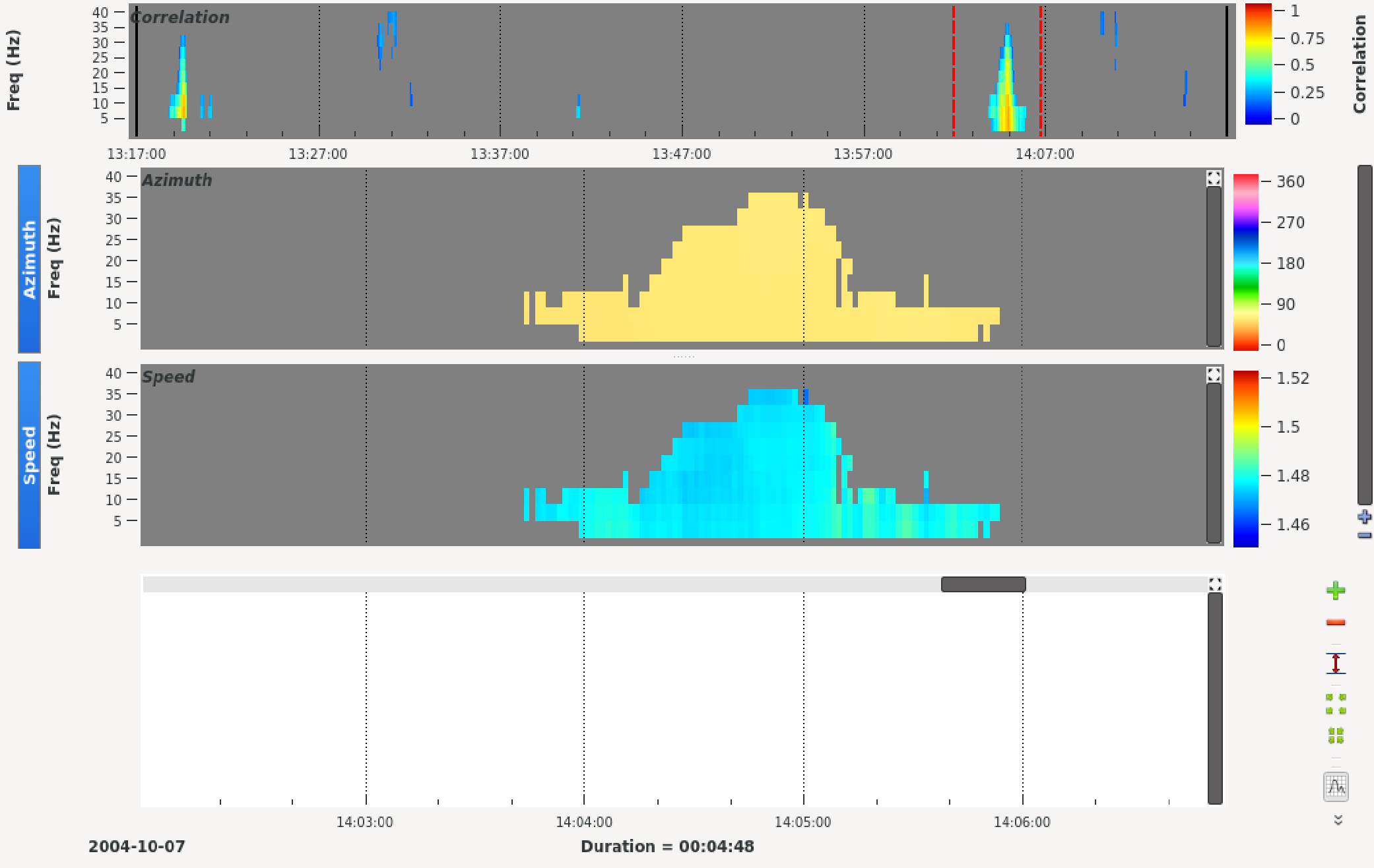}
  \caption{The DTK-GPMCC measurement of the possible reflected signal for the Oct 7, 2004 fireball recorded at H08S. The top panel shows signal correlation in the full 1h search window - red vertical lines denote the shorter timeframe of the middle two windows. Middle window - pixel by pixel backazmiuth for the signal. Bottom window - pixel by pixel estimate of the trace speed of the signal.}
  \label{2004_spectra}
\end{figure*}

\begin{figure*}[!ht]
  \includegraphics[width=\linewidth]{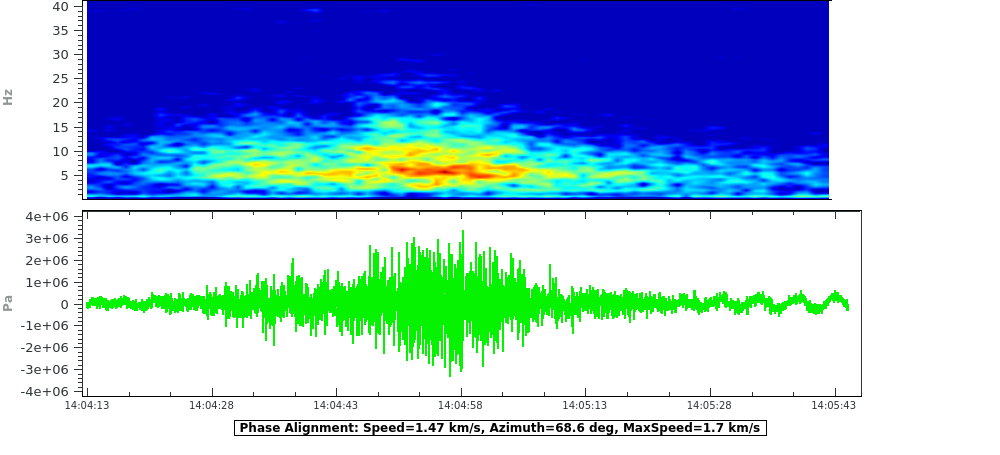}
  \caption{The stacked waveform and spectrogram of the possible reflected hydroacoustic signal associated with the Oct 7, 2004 fireball at H08S.}
  \label{2004_wave}
\end{figure*}

Station H08S showed a clear, high SNR signal in unfiltered data at all three array elements in the later part of the 1h search window as shown in Figures \ref{fig:2004_H08}, \ref{2004_spectra}, \ref{2004_wave}. This arrival is impulsive, very clear, has high signal-to-noise and is in fact the most prominent within a 5-hour window centered on the meteor-related signal at H08S. 

However, this signal appeared at 14:04:57 UTC from an azimuth of 69 degrees, instead of the expected direct arrival time of 13:38 UTC and azimuth of 182, approximately 110 degrees from the expected great circle direct path from the fireball. Given the large energy of the fireball, this suggested it was worth exploring if the direct path from source to receiver was obscured and that the late, large signal was a reflection. Table \ref{2004_data_table} gives the expected direct arrival time (not observed) and the timing of the later observed arrivals which might be reflections.

\begin{figure*}
  \includegraphics[width=\linewidth]{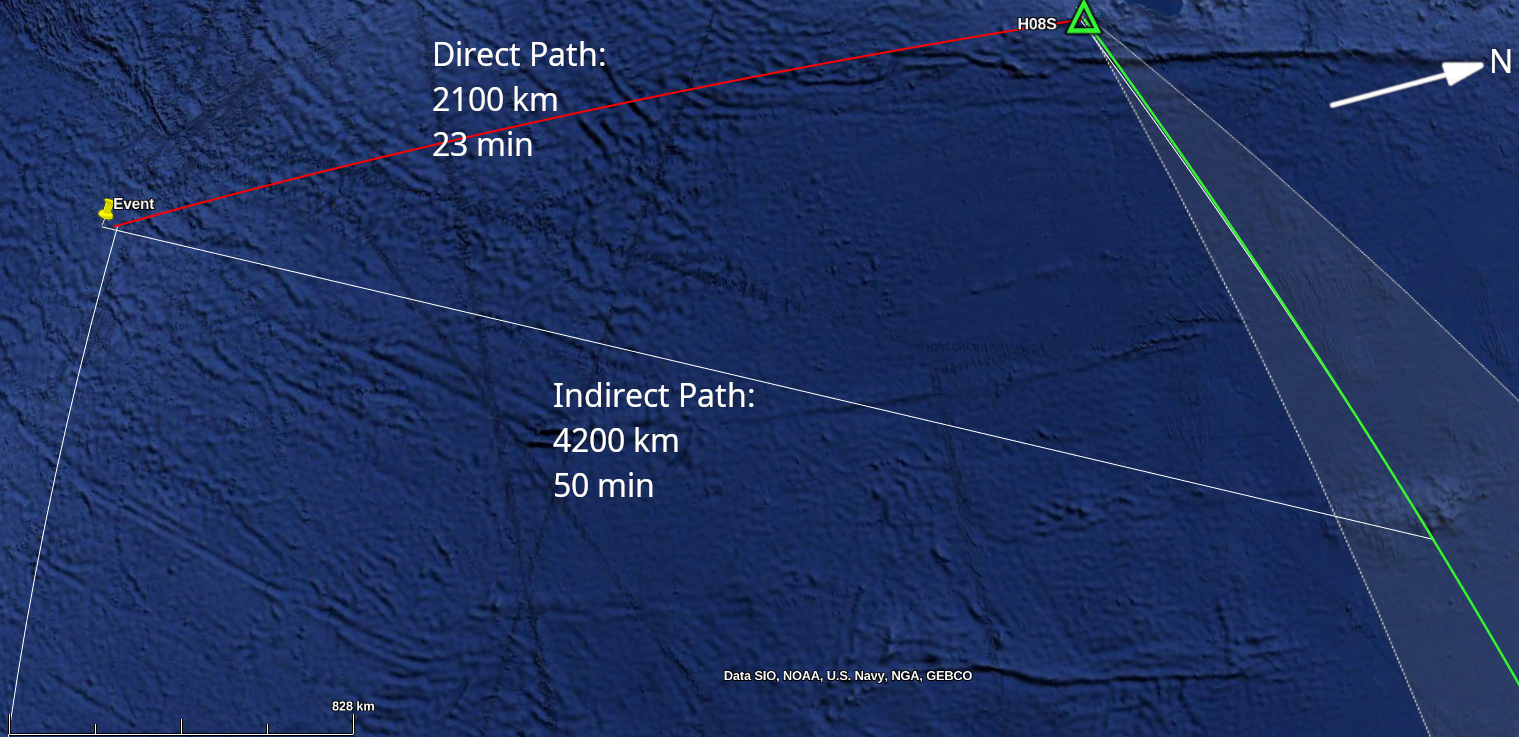}
  \caption{An overview of the Oct 7, 2004 fireball-hydroacoustic event, showing the path of both the (unobserved) direct arrival at H08S in red, and the approximate path of the observed indirect arrival in white together with the expected travel times. The observed signal backazimuth and its uncertainty are shown by the green line and white envelope.}
  \label{2004_overview}
\end{figure*}

We manually examined the local bathymetry for possible reflectors, and found one at approximately 3.44S, 83.44E, from the Afanasij Nikitin Seamount, which has a minimum depth of 1800 m below the water surface, marginally approaching the 1500m maximum SOFAR depth in this area predicted by \citet{Chu_Fan_2024}. Figure \ref{2004_overview} shows the proposed reflection path, as well as the unobserved direct arrival. \par

\clearpage


\begin{figure*}
  \includegraphics[width=\linewidth]{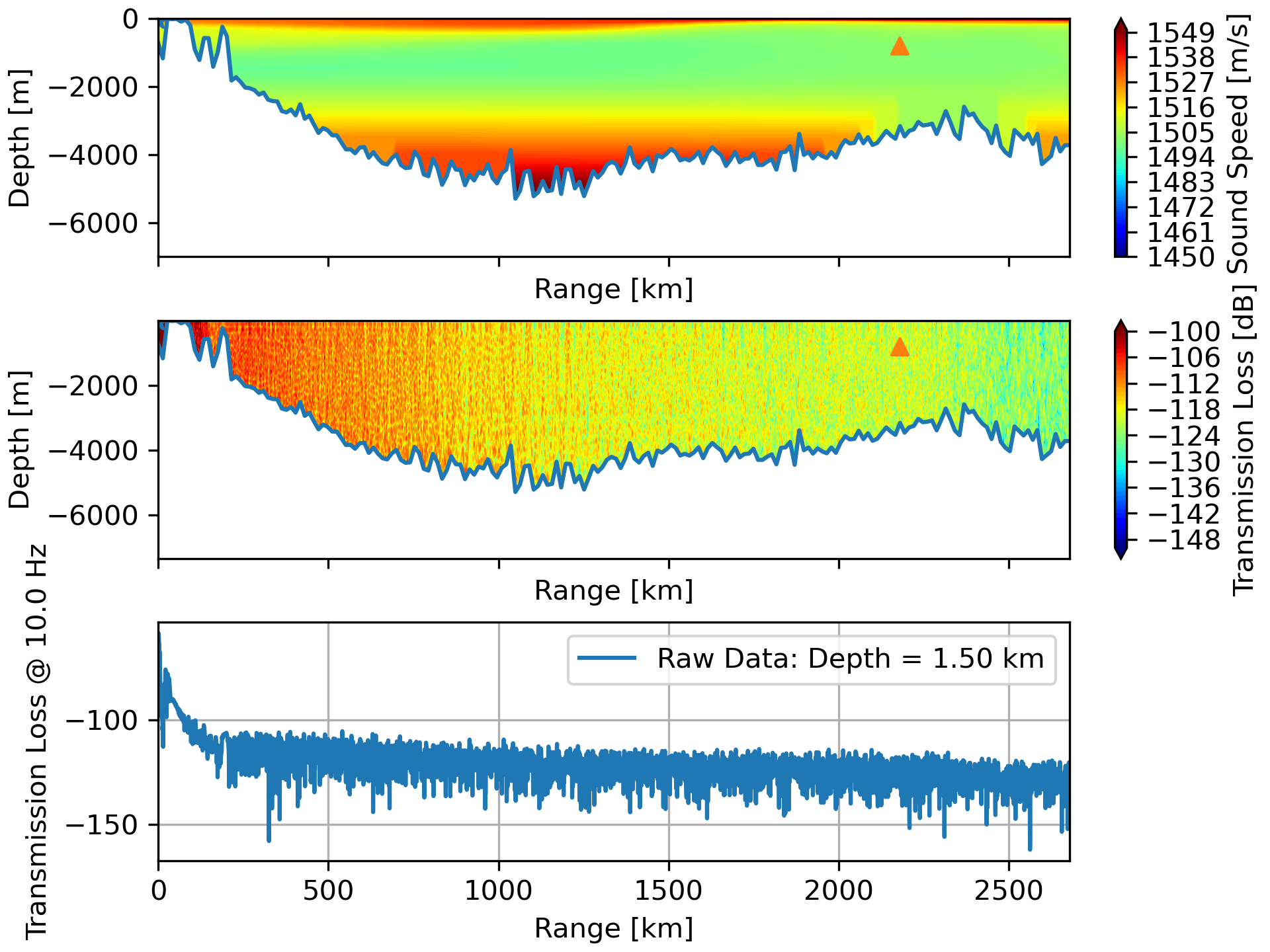}
  \caption{The modeled transmission loss along the direct raypath from the Oct 7, 2004 fireball sub-terminal point (at a range of 0 km)  to the H08S array (shown as orange triangle) using PyRAM.  The top plot shows the bathymetric profile as a function of range along the direct great-circle path from the source to the receiver together with the local sound speed (color bar) in m/s. The middle plot shows the transmission from a surface source as a function of range and depth with the color bar representing transmission loss in dB. The bottom plot is an estimate of the transmission loss at a frequency of 10 Hz along the SOFAR channel, which we take to be centered at 1.5km depth.}
  \label{2004_direct_path}
\end{figure*}

\begin{figure*}
  \includegraphics[width=\linewidth]{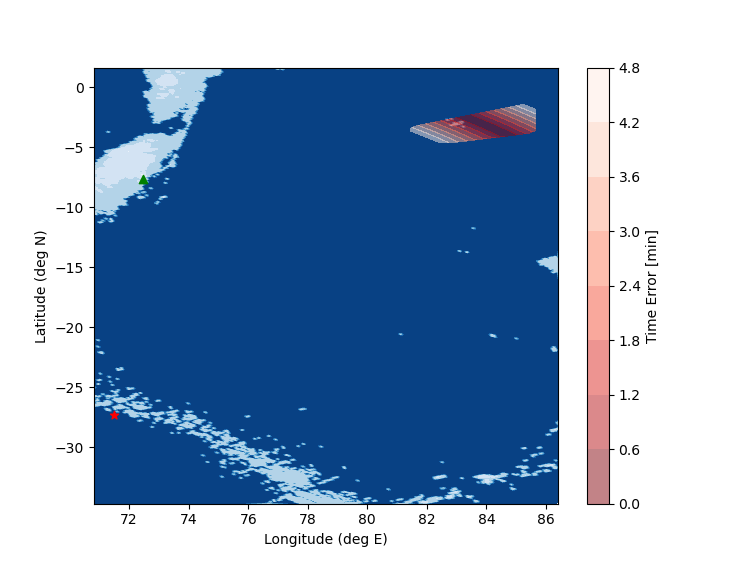}
  \caption{A bathymetric map covering the region near the impact location of the Oct 7 2004 fireball (red star) and the IMS hydroacoustic station H08S (green triangle). Regions where the ocean floor is more than 3 km deep are shown in dark blue while areas where topography reaches to depths less than 3 km are light blue. The red contour near the promontory region in the top right shows the simulated reflection points which align within 5 degrees of the signal backazimuth observed at H08S. The difference in timing between the observed time delay and modeled are shown by the colorbar. As the  lowest temporal difference is co-located with significant and isolated bathymetric features, this is a likely reflection path for the fireball hydroacoustic signal.}
  \label{2004_bathymap}
\end{figure*}

\begin{figure*}
  \includegraphics[width=\linewidth]{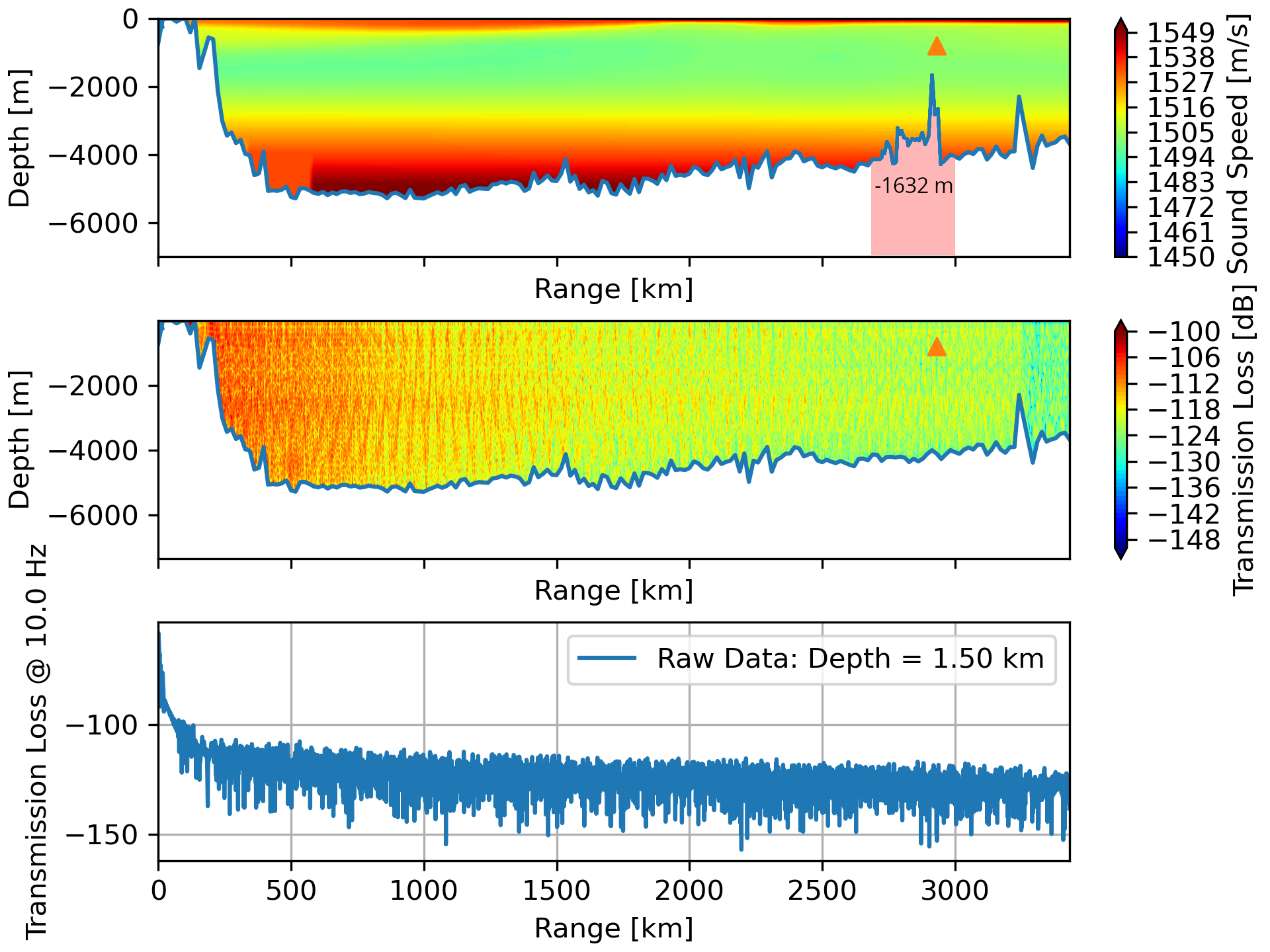}
  \caption{The PyRAM output of the indirect path for the Oct 7, 2004 fireball, from the meteor source to the reflector (shown as the orange triangle in the top plot). The top plot shows the bathymetric profile as a function of range along the reflector great-circle path from the source to the proposed reflector together with the local sound speed (color bar) in m/s. The middle plot shows the transmission from a surface source as a function of range and depth with the color bar showing the transmission loss in dB. The bottom plot is an estimate of the transmission loss at a frequency of 10 Hz along the SOFAR channel, which we take to be centered at 1.5km depth. 
  Here we observe approximately the same transmission loss as in Figure \ref{2004_direct_path}, which indicates that an acoustic ray would be blocked as with the direct arrival. However, due to the shape of the bathymetry as shown in Figure \ref{2004_bathymap}, a ray could potentially follow an indirect path as shown in Figure \ref{2004_overview} with sufficient residual energy to scatter and be observed at H08S, allowing for a path with multiple reflections through the bathymetric feature close to the source. Due to the resolution of our bathymetry data, we were able to show the lower parts of the underwater hill near the reflector, but were unable to show the peak of the mountain available in synthesized datasets such as what is used by Google Earth, as it was interpolated as the rest of the hill. Here, we have superimposed the bathymetry of the mountain from Google Earth as a red region, with the highest elevation recorded as -1632 m. For reference, the second subplot shows the unaltered bathymetry.}
  \label{2004_indirect_path1}
\end{figure*}

We searched for possible undersea reflectors using the observed timing and backazimuth as constraints and superimpose these on the bathymetry map. Figure \ref{2004_bathymap} shows that a propagation path where reflection occurs at the seamount, consistent with the observed backazimuth of $68.96^{\circ}$, and with the arrival time of 14:04:57 UTC within tolerances.

\begin{figure*}
  \includegraphics[width=\linewidth]{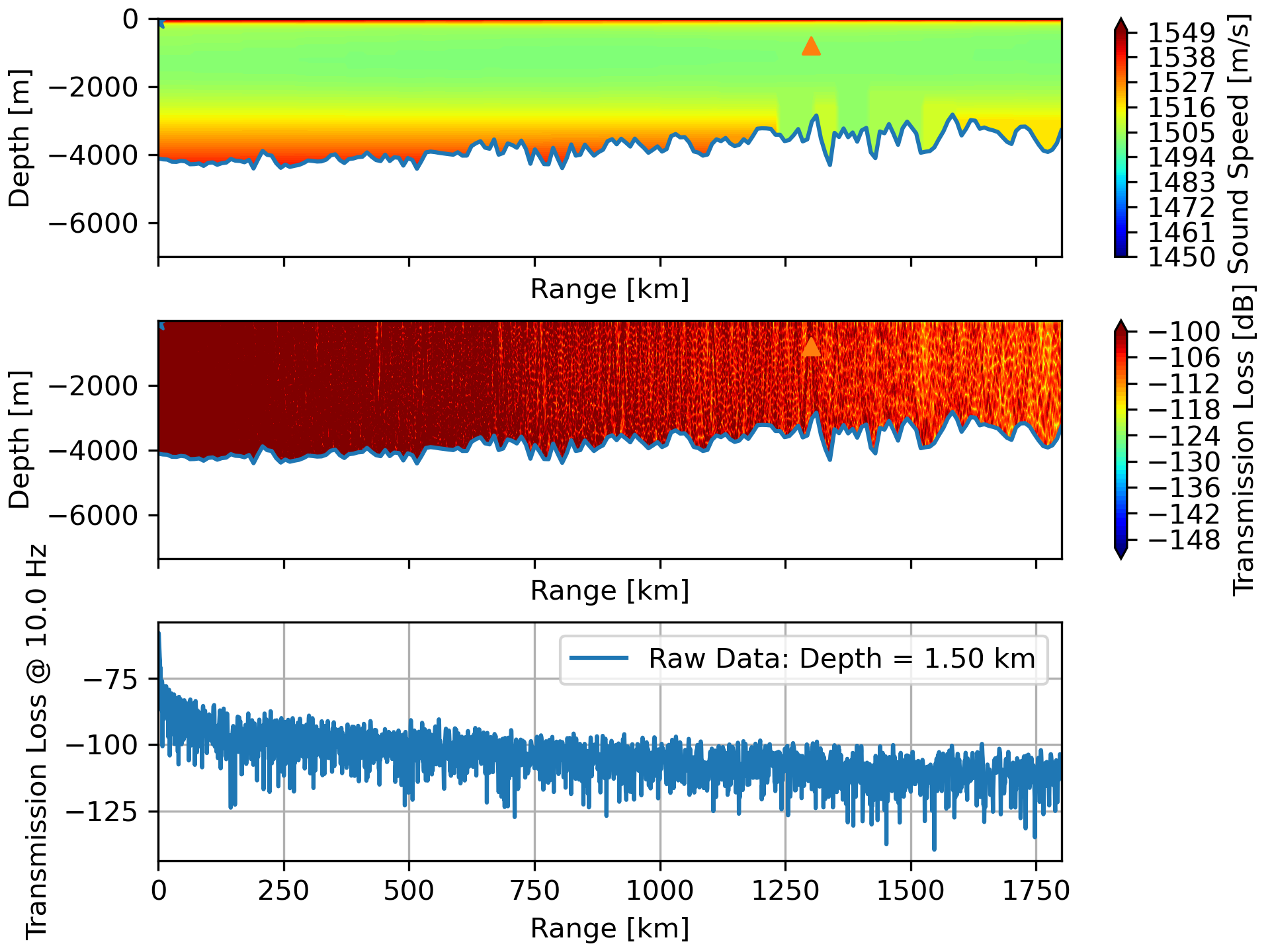}
  \caption{The PyRAM output for the Oct 7, 2004 fireball of the indirect path, from the reflector to the H08S station (shown here as the orange triangle). This is a continuation of the ray-path following reflection from the source as summarized in Figure \ref{2004_indirect_path1}. Here we have shown the unaltered bathymetry, but the reflector is shown in Figure \ref{2004_indirect_path1} for reference.}
  \label{2004_indirect_path2}
\end{figure*}

Figures \ref{2004_indirect_path1} and \ref{2004_indirect_path2} show the transmission losses between source and reflector, and reflector and receiver, respectively. The transmission loss along the direct path (Figure \ref{2004_direct_path}) and in the direction from the source to the potential reflector are similar. 


Unfortunately, at the time of the fireball the H04N and H04S arrays (with a range of 2700 km away, and heading $215^{\circ}$, from the event) only had two active elements precluding backazimuth measurements. Additionally, while we found a potential signal on H08N it was much weaker than H08S, possibly due to differences in the local bathymetry near the stations.

The waveform shape of this strong arrival is very similar to the lens-shaped coda typical of most T-phase arrivals \citep{Williams_2006}. Similarly, T-phase spectra normally show a peak in the 2-8 Hz range \citep{Schwardt2022}, very similar to our signal.

The simplest explanation of the foregoing facts is that this reflected signal is in fact unrelated to the fireball. This is supported by the nominal propagation modeling, which suggests comparable signal strength should be observed as a direct arrival but the latter is unobserved and by the lack of any detections at H01 or H04 both of which have clear propagation paths to the fireball.

\clearpage

\end{document}